\documentclass[12pt]{article}
\usepackage{latexsym}
\usepackage{epsfig,amssymb,euscript,graphicx}
\usepackage{amsmath}
\numberwithin{equation}{section}  

\newsavebox{\ns}
\newsavebox{\dbrane}
\newsavebox{\dbshort}

\def\be{\begin{equation}}
\def\ee{\end{equation}}
\def\bea{\begin{eqnarray}}
\def\eea{\end{eqnarray}}

\def\Dslash{\,\,{\raise.15ex\hbox{/}\mkern-12mu D}}
\def\Dbarslash{\,\,{\raise.15ex\hbox{/}\mkern-12mu {\bar D}}}
\def\delslash{\,\,{\raise.15ex\hbox{/}\mkern-9mu \partial}}
\def\delbarslash{\,\,{\raise.15ex\hbox{/}\mkern-9mu {\bar\partial}}}
\def\pslash{\,\,{\raise.15ex\hbox{/}\mkern-9mu p}}
\def\calDslash{\,\,{\raise.15ex\hbox{/}\mkern-12mu {\cal D}}}

\newcommand\R{\mathbb{R}}
\newcommand\Z{\mathbb{Z}}

\newcommand{\de}{\partial}
\newcommand{\nn}{\nonumber}
\newcommand{\eps}{\epsilon}
\newcommand{\vol}{\mathrm{vol}}

\newcommand{\e}{\mathrm{e}}

\newcommand{\omehol}{\Omega_\mathrm{hol}}

\def\nref#1{(\ref{#1})}
\def\MyUpsilon{{S^3}}
\def\horizon{{brane }}

\begin{document}
\begin{titlepage}
\begin{center}

\vspace*{1.5cm}

{\Large \bf  The unwarped,   resolved, deformed conifold:\\[.4mm]
 fivebranes and the baryonic branch of\\[2mm] the Klebanov-Strassler theory}

\vskip 15mm
Juan Maldacena$^1$ and Dario Martelli$^2$\\
\vskip 10mm

{$^1$\em Institute for Advanced Study\\
1 Einstein Drive, Princeton, NJ 08540, U.S.A.}\\

\vskip 1.2 cm

{$^2$ \em  Swansea University,\\
Singleton Park, Swansea, SA2 8PP, U.K.}\\

\vskip 3.5cm

\end{center}

\begin{abstract}
\noindent
We study a gravity solution corresponding to fivebranes wrapped on the $S^2$
of the resolved conifold. By changing a parameter the solution continuously interpolates
between the deformed conifold with flux and the resolved conifold with branes.
  Therefore, it displays a geometric transition, purely in the supergravity context.
The solution is a simple example of torsional geometry and may be thought
of as a non-K\"ahler analog of the conifold.
By U-duality transformations we can add D3 brane charge and recover the solution in the form
originally derived by Butti et al.  This describes the baryonic branch of the
 Klebanov-Strassler theory.
Far along the baryonic branch the field theory gives rise to a fuzzy two-sphere. This corresponds
to the D5 branes wrapping the two-sphere of the resolved conifold in the gravity solution.
\end{abstract}

\vfill

\end{titlepage}

\section{Introduction}

The conifold \cite{Candelas:1989js}
is a very simple non compact Calabi-Yau geometry which has given us important lessons
about the behavior of string theory. In particular, understanding the transition between
the resolved and the deformed conifold was very important \cite{Andycon}.
The conifold geometry was also important for constructing the Klebanov-Strassler geometry \cite{KS},
which is
dual to a four dimensional  ${\cal N}=1$ supersymmetric field theory displaying confinement.
This field theory arises when one wraps fivebranes and antifivebranes on the conifold and one
takes the near \horizon limit\footnote{
This is sometimes called the ``near horizon'' limit. However, since the solutions in this paper
 have no horizon, it is more appropriate
to call this a ``near brane'' limit.}.
This theory can spontaneously break a $U(1)_B$ baryonic symmetry.  The geometries corresponding to
arbitrary values of the corresponding VEVs  for  the baryonic operators
were constructed by Butti et al. in  \cite{Butti} and
further studied in \cite{DKS,BDK} and several other papers.

Here we further analyze the solution in \cite{Butti}. We point out that the solution in \cite{Butti} is related
to a simpler solution which   corresponds to fivebranes wrapping the $S^2$ of the resolved conifold and no
extra D3 brane charge.
The solution we discuss is not new, it is a limit of \cite{Butti}, and was also  discussed in \cite{nunezetal}.

  Another supergravity solution
 with  ${\cal N}=1$ supersymmetry    is the
Chamseddine-Volkov/Maldacena-Nu{\~n}ez solution \cite{volkovone,volkovtwo,MN}.
This is the near \horizon region for  fivebranes wrapped on the two-sphere of the resolved conifold.
Indeed, we will see that both the CV-MN geometry and the KS solution arise as limits of this more general solution.

This solution with fivebranes wrapping the $S^2$ of the resolved conifold displays
very clearly the geometric transition
described in \cite{vafageometric}. In fact,
 the solution depends on a non-trivial parameter which can roughly be viewed as the size of the two-sphere that the brane is wrapping.
When this size is very large, the solution looks like the
resolved conifold with branes and when it is very small the solution becomes
the deformed conifold with three form flux on $S^3$.
The geometry is always smooth and has the topology and also the complex structure
of the deformed conifold. However, the metric is not Ricci-flat.

In fact, if we consider only NS-5 branes, this solution is a very simple example of a non-K\"ahler, or torsional, manifold
\cite{Strominger:1986uh,Hull:1986iu,Lopes Cardoso:2002hd,GMW,becker,Fu:2006vj,yaukahler,kahlernoncompact,Dasgupta:1999ss,Becker:2009df}.
It is a solution with three form flux $H_3$ which preserves ${\cal N}=1$ supersymmetry. Thus, the solution we consider could
be viewed as the non-K\"ahler analog of the conifold. Namely, it is a simple and very symmetric geometry which could be
describing smaller regions of a bigger six dimensional geometry.
As was discussed in \cite{vafageometric}, the addition of flux or branes smoothly connects
the deformed and the resolved
conifolds. Here we are stressing, that this transition can be seen rather vividly purely in the gravity context.
This is in contrast to the   similar transition involving D6 branes on the $S^3$ of the deformed
 conifold \cite{AMV} which requires
some non-geometric insight.

We then argue that starting from this solution we can construct a solution which also has D3 brane charge.
This is done via a U-duality symmetry of the classical gravity equations.
This configuration corresponds to a cascading theory, as in \cite{KS}, but where the cascade stops and merges into
an ordinary conifold with constant four dimensional  warp factor.
A further scaling limit of this solution gives the usual Klebanov-Strassler solution \cite{KS} where the cascade goes on forever.
The resulting geometry represents the baryonic branch of the quiver field theory \cite{Butti,DKS}.

This perspective on the solution is useful for understanding how the geometry behaves for large values
of the VEV of the baryonic operators. For large values of the baryonic operators the solution looks like
fivebranes wrapping the $S^2$ of the resolved conifold. The fivebranes have some dissolved D3 branes.
 This leads to a non-commutative theory on the fivebrane worldvolume.

In fact, one can see this $S^2$ emerging already in a weakly coupled version of the quiver field theory.
We show that the baryonic VEVs written in the field theory analysis of \cite{DKS} can be viewed as a \emph{fuzzy two-sphere}. The spectrum of field theory excitations match the Kaluza-Klein modes on the fuzzy two-sphere. The sphere becomes less and less fuzzy as we increase the VEVs of the baryonic operators.
Thus, for large baryon VEVs we recover the picture of a fivebrane
wrapping the $S^2$ of a resolved conifold in a way that matches rather precisely  the gravity description.

The emergence of the fivebrane theory from the KS theory may be also understood from a dielectric effect,
in a fashion rather analogous to \cite{Polchinski:2000uf}.
It was demonstrated in \cite{Andrews:2006aw} that the fivebrane theory on the fuzzy
two-sphere emerges from   the mass-deformed ${\cal N}=1^*$ theory. Our derivation here
from the baryonic branch of the KS theory displays some novel features. In particular,
in our construction the fuzzy two-sphere emerges from bifundamentals in
a quiver matrix theory. Recently a very
closely related
construction was discussed in \cite{mark,Nastase:2009ny} for vacua of the mass-deformed ABJM theory \cite{abjm}.

We expect that a similar picture would hold for other field theories
coming from D3 branes on singularities plus fractional branes.
Far along the baryonic branch we expect to see resolved geometries with fractional branes.

\section{Fivebranes on the two-sphere of the resolved conifold}
\label{section_one}

\begin{figure}[ht!]
\epsfxsize = 9cm
\centerline{\epsfbox{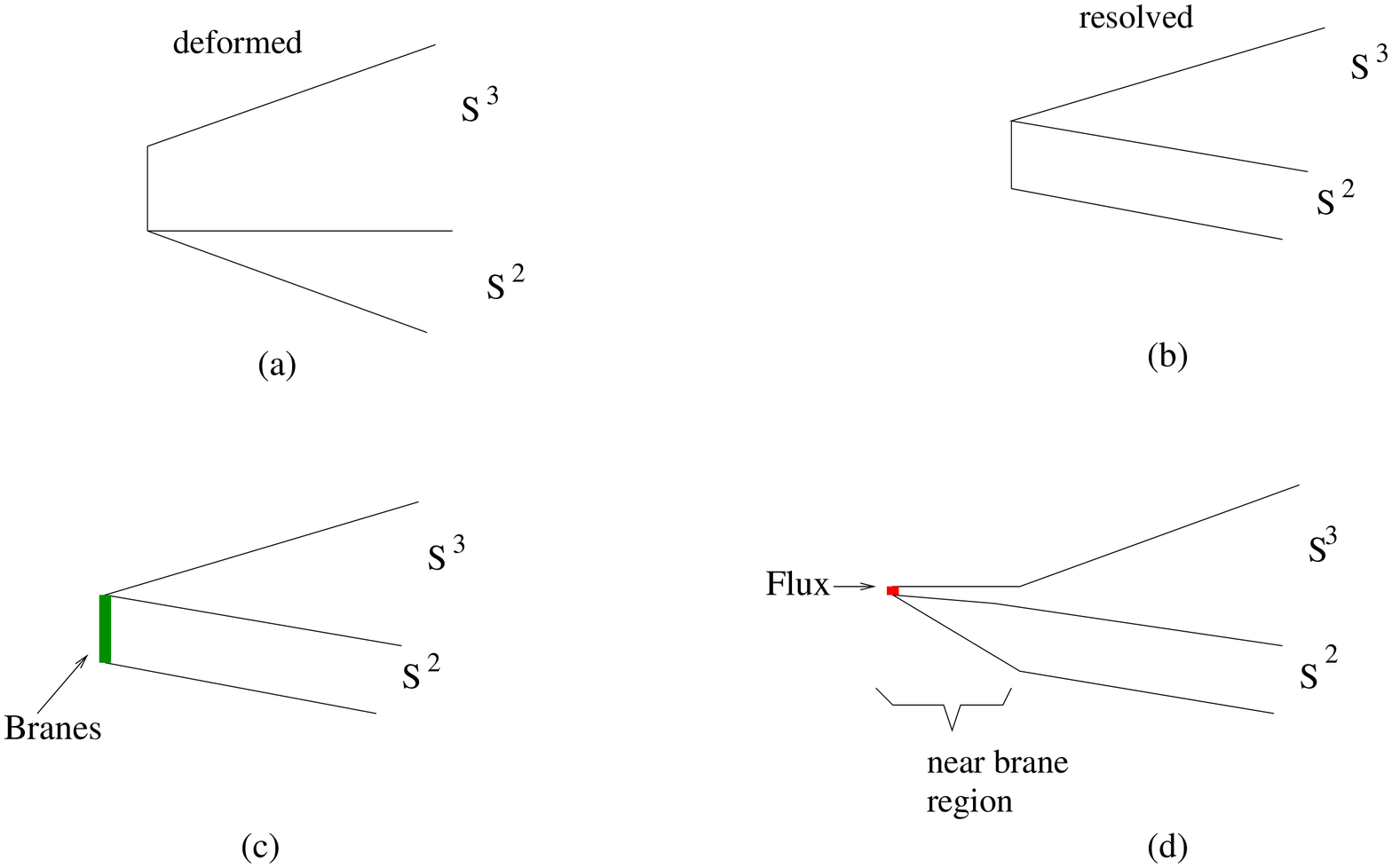}}
\caption{(a) A picture of the deformed conifold. The $S^2$ shrinks but the $S^3$ does not. (b) The resolved
conifold, the $S^2$ does not shrink but the $S^3$ shrinks. In (c) we add five branes wrapping the $S^2$ of
the resolved conifold of picture (b). (d) Backreacted geometry. The branes are replaced by geometry and fluxes. The
end result is a geometry topologically similar to that of the deformed conifold with flux on the $S^3$.
The near \horizon region is the CV-MN solution \cite{volkovone,volkovtwo,MN}.}
\label{conifold}
\end{figure}

In this section we discuss the gravity solution that is associated to fivebranes wrapping
the $S^2$ of the resolved
conifold. In other words, it is the solution that takes into account the backreaction of the branes on the geometry.
It is a smooth solution with flux. The final topology of the solution is  that of the
deformed conifold, in the sense that there is an $S^3$ which is not shrinking, see figure \ref{conifold}.
This solution is a limit of the one found in \cite{Butti}, and was also written in \cite{nunezetal}.
We will discuss some of its
properties in some detail. We will also see that this solution can be used to construct
the full solution in \cite{Butti}.

In order to write the solution we will use the NS language. Namely, we consider NS-5 branes wrapping the
$S^2$ of the resolved conifold. The solution corresponding to D5 branes can be easily found by performing an S-duality.
The NS solution can also be interpreted as a solution of type IIA supergravity or even the heterotic string theory
(or type I supergravity)\footnote{ For the heterotic application we can embed the spin connection into
the gauge group in order to cancel the $Tr[R \wedge R ] - Tr[F\wedge F]$ term. Otherwise, it should also
be taken into account and it would lead to a modification of the solution. This modification is
 small in the limit of a large number of branes.}.

The solution is a  simple example of a non-K\"ahler (or torsional) geometry involving $H_3$ flux,
 originally studied by Strominger
\cite{Strominger:1986uh} and Hull \cite{Hull:1986iu}.
See also \cite{Lopes Cardoso:2002hd,GMW,yaukahler} for more recent discussions of these geometries.
The solution contains the NS three form, the metric and
the dilaton. These are non-trivial only in six of the ten dimensions. Other non-compact non-K\"ahler geometries
were discussed in \cite{kahlernoncompact}.

\begin{figure}[ht]
  \hfill
  \begin{minipage}[t]{0.49\textwidth}
    \begin{center}
    \epsfxsize=7.2cm
    \epsfbox{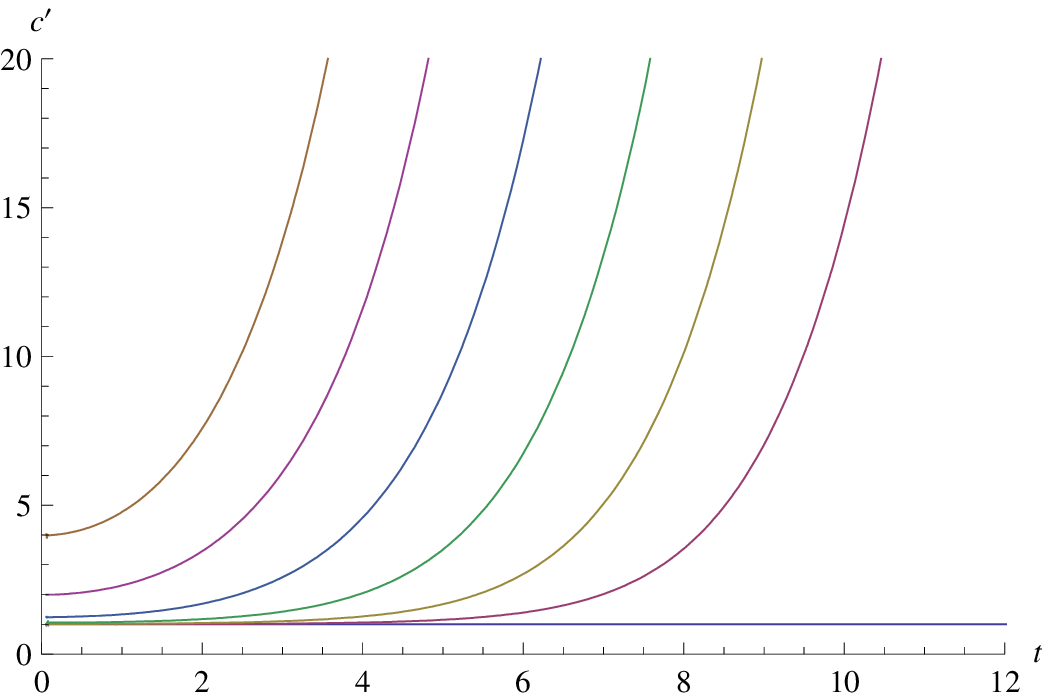}
    \end{center}
  \end{minipage}
  \hfill
  \begin{minipage}[t]{.49\textwidth}
    \begin{center}
     \epsfxsize=7.2cm
    \epsfbox{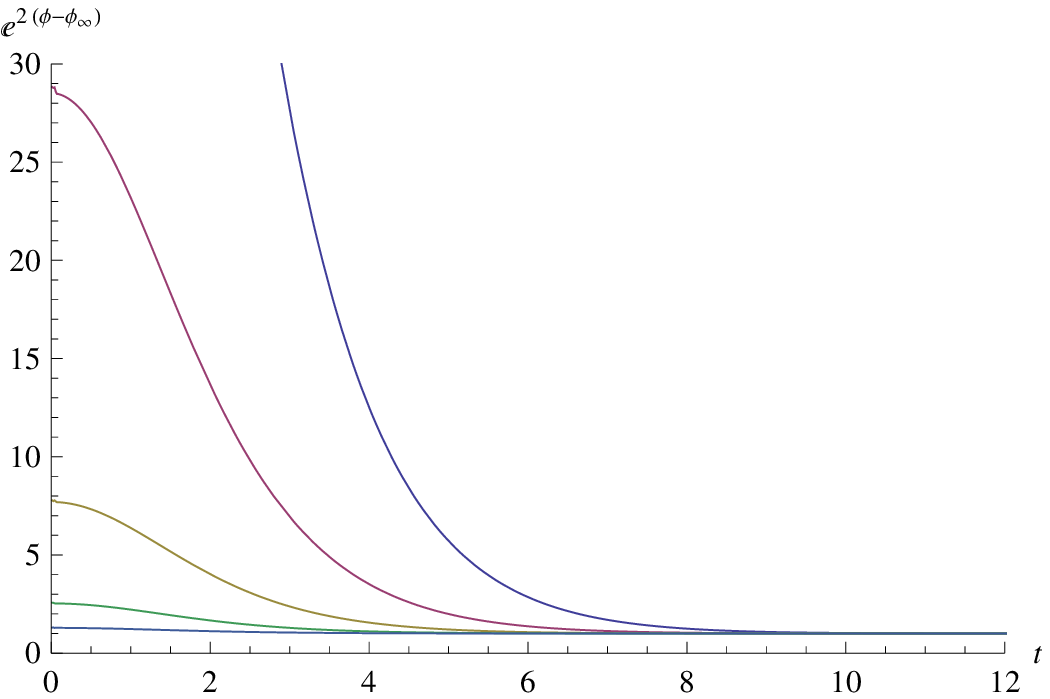}
     \end{center}
  \end{minipage}
\caption{Plots of the solutions for some values of $\gamma^2$.
On the left hand side: plots of $c'$ for $\gamma^2=1.01,1.02,1.06,1.25,2,4$. The bottom constant one is the CV-MN value, $\gamma^2=1$.
On the right hand side: plot of $e^{2(\phi-\phi_\infty)}$ for $\gamma^2=1.01,1.02,1.06,1.25,2$.
The CV-MN profile is not plotted since the dilaton does not asymptote to a constant.}
\hfill
  \label{twoplots}
\end{figure}

The solution is \cite{Butti,nunezetal}
  \bea \label{ansatz}
ds^2_{str} &=& dx_{3+1}^2 + {  \alpha' M\over 4 }  ds^2_6
\\
ds^2_6 &=&  c' ( dt^2 + (\epsilon_3 + A_3)^2) +
{ c \over \tanh t} ( \epsilon_1^2 + \epsilon_2^2 + e_1^2 + e_2^2)  +
2 { c \over \sinh t} ( \epsilon_1 e_1 + \epsilon_2 e_2)  \notag
\\ && ~~ + \left( { t \over \tanh t } - 1 \right)
( \epsilon_1^2 + \epsilon_2^2 - e_1^2 - e_2^2)\label{themetric}
\\
e^{ 2 \phi} &=& e^{ 2 \phi_0} { f^{1/2} c '  \over \sinh^2 t  }~, \label{dilatonvalue}
~~~~~~~~~~~~~H_3 ~=~{  \alpha' M \over 4 } w_3~ ,~~
\\
w_3 &=&   ( \epsilon_3 + A_3)\wedge \left[ ( \epsilon_1 \wedge \epsilon_2 + e_1 \wedge e_2 ) + {t \over \sinh t}
( \epsilon_1 \wedge e_2 + e_1 \wedge  \epsilon_2 )\right]  \notag \\
&&  ~~~~ + {(t \coth t -1 )\over \sinh t}
dt \wedge ( \epsilon_1 \wedge e_1 + \epsilon_2 \wedge e_2 )
\eea
 where
 \bea
 e_1 &=& d  \theta_1 ~,~~~~~~e_2 = - \sin \theta_1 d\phi_1 ~,~~~~~~~~~~A_3 = \cos \theta_1 d\phi_1~,\notag
 \\
 \epsilon_1 + i \epsilon_2 &=& e^{ - i \psi} ( d\theta_2 + i \sin \theta_2 d \phi_2 ) ~,~~~~~~~
 \epsilon_3 = d \psi + \cos \theta_2 d \phi_2~.
 \eea
The $SU(2)$ left-invariant one-forms $\epsilon_i$ obey $d\epsilon_1 = - \epsilon_2 \wedge \epsilon_3$ and cyclic permutations.
The functions $c(t)$ and $f(t)$ appearing in \nref{ansatz} obey the equations
 \bea
 f ' &=& 4 \sinh^2 t \, c \label{equationf}
 \\
 c' &=& { 1 \over f } [ c^2 \sinh^2 t - (t \cosh t -  \sinh t )^2 ]  \label{equationc}
 \eea
where the primes denote derivatives with respect to $t$. The range of $t$ is between zero and infinity.
 We will be interested in solutions to these equations with the following boundary conditions
for small  and large $t$
 \bea \label{smallt}
  && c = \gamma^2 t + \cdots , ~~~~~~~~~~~~~~~~~~ f =  t^4 \gamma^2  + \cdots~,   ~~~~~~~~~~~~~~~~{\rm for} ~~~t \to 0
 \\ \label{larget}
 && c ={ 1 \over 6 } e^{ 2 (t - t_{\infty} ) \over 3 } + \cdots
   ~,~~~~~~~~~ f =
  { 1 \over 16 }  e^{ 2 t_{\infty} }  e^{ 8 (t - t_{\infty} ) \over 3 } + \cdots ~,~~~~{\rm for}~~~ t\to \infty
  \\
  && U   \equiv  12e^{ 2 t_\infty \over 3 }
   \eea
  where the dots indicate higher order terms. $\gamma^2$ and $t_\infty$ are parameters describing a family of
  solutions. We have also related the parameter $t_\infty$ to the parameter $U$ introduced previously in the literature
  \cite{DKS}.
 In general, we can only solve the equations numerically. The solutions have the property that
 both $c$ and $f$ are monotonically increasing and thus $c'$ is positive.
We plot some representative  solutions in figure \ref{twoplots}.
In appendix A  we comment further on the structure of these equations. The solutions interpolate between the conifold
 at $t \to \infty$ and the $t =0$ region where there is an $S^3$ which does not shrink and has flux $M$ for
 the $H_3$ field, see figure \ref{conifold}(d).
There is a one parameter family of solutions since the differential equations relate $\gamma^2$ to
  $t_\infty$. The parameter $\gamma^2$ should be larger
  than one, and $t_{\infty}$ can have any real value. The dilaton is a maximum at $t=0$ and it decreases when
  we go large values of $t$, achieving a constant value $\phi_{\infty}$ asymptotically as $t\to \infty$.

This way of writing the metric and the equations makes manifest the $\Z_2$ symmetry corresponding to making a
flop transition of the conifold (before we wrap the branes) and then
 wrapping an antibrane on the flopped two cycle. Explicitly, this
symmetry is $ c, f, M \to -c,-f,-M$. This is a symmetry of the equations, but not of the solutions.

The solution has an $SU(2) \times SU(2)$ global symmetry.

 In addition, the full configuration has a second  parameter which corresponds to  an overall shift of the dilaton,
 denoted by $\phi_0$. From the gravity point of view this is a rather trivial parameter.
Finally, there is also an overall size parameter $M$. This appears also in $H_3$ where it gives the flux of three
  form on the $S^3$. It is  thus quantized and the integer $M$ can be viewed as number of
fivebranes that we are wrapping.

\begin{figure}[ht!]
\epsfxsize = 9cm
\centerline{\epsfbox{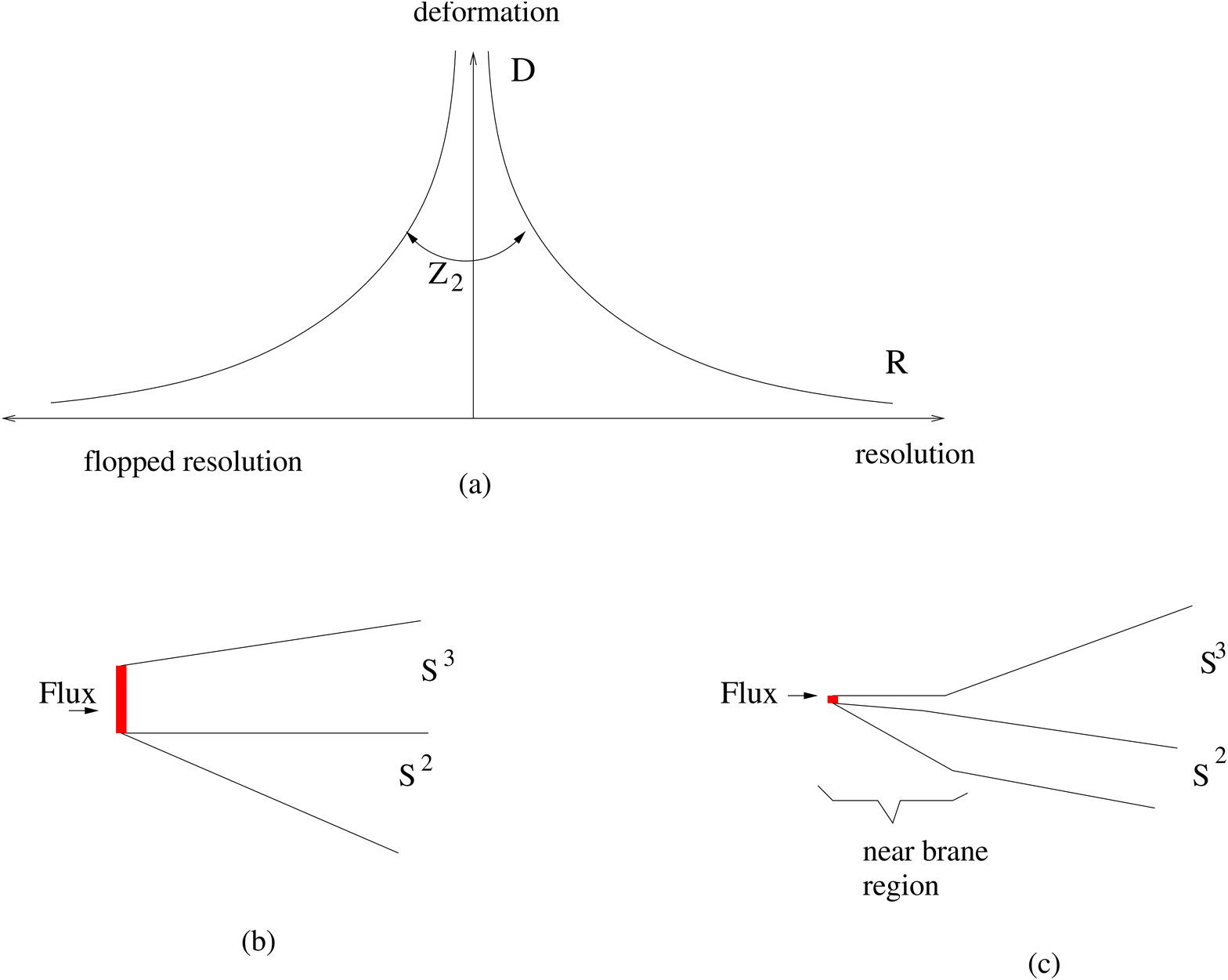}}
\caption{(a) The moduli space of the conifold with no flux or branes has two branches, denoted here by the
vertical and horizontal axes. One is a deformation and the other is the resolution, which has two sides differing by flop
transition. When we add flux
we have a one parameter family that interpolates continuously between a deformed conifold with flux in region $D$ and
a resolved conifold with branes in region $R$. A $\Z_2$ symmetry relates this to another branch that joins the deformed conifold
with the flopped resolved conifold. (b) The solution looks like the deformed conifold with flux in
region $D$ of (a). (c) In region $R$ of (a) the solution looks like the resolved conifold with some
branes, where the branes have been replaced by their  near \horizon  geometry. In all cases the topology (but
not the geometry)  is that of the deformed conifold. }
\label{regions}
\end{figure}

  Let us discuss in more detail the dependence of the solution on the nontrivial parameter $\gamma^2$ or $t_\infty$.
 At the level of gravity solutions we can view this parameter as the size of the $S^3$ at the origin.
 Namely, the  $S^3$ at the origin has radius squared equal to $r^2_{S^3} = \alpha' M   \gamma^2 $.
 However, from the quantum gravity perspective,
    it is more convenient to define the parameter
 at large distances, where the geometry is more rigid,
 by stating how the metric deviates from the conifold metric.
 Naively one would think that the parameter should simply be the size of the $S^2$ of the resolved
 conifold that the brane is wrapping.
 More explicitly,  the metric of the resolved conifold is
 \cite{PZT}
\bea
d s^2_{RC} \, &=&
\, \frac{1}{\kappa (\rho)} d \rho^2 + \frac{\rho^2}{6} (e_1^2 +e_2^2) + \left(\alpha^2+\frac{\rho^2}{6}
\right) (\eps_1^2+\eps_2^2)+ \frac{\rho^2}{9}\kappa (\rho) (\epsilon_3 +A_3)^2
\notag
\\
&&
 ~~{ \rm
where }~~~~
\kappa (\rho)=  \frac{9\alpha^2 +\rho^2}{6 \alpha^2 + \rho^2}~. \label{RCmetric}
\eea
This is the metric before adding any branes or fluxes.
As $\rho\to 0 $ the metric approaches a two-sphere  $S^2$, of radius $\alpha$,    and four normal
directions that are fibered over the two-sphere.
Looking at the metric far away, at large $\rho$, we can view the
 resolution parameter $\alpha^2$ as the difference
between the coefficients of  $\epsilon_1^2$ and $e_1^2$, for example. This is a non-normalizable mode and we
can view it as a parameter that we can control
 from infinity. This parameter  eventually sets the size of the $S^2$ at
$\rho =0$, but is, in principle, defined at large $\rho$.

One would   naively expect  that the full geometry  \nref{ansatz}   asymptotes to the large $\rho$
form of \nref{RCmetric}.
On the other hand, returning to \nref{ansatz}  and inserting \nref{larget},
 we see that the large $t$ asymptotic form has an effective $\alpha^2$ which is linear in $t$ for large $t$,
 coming from
  the term of the form $({ t \over \tanh t} -1)$ in \nref{ansatz}.
 Thus, we see that we cannot fix this parameter at infinity,
 it ``runs'' with the distance. This is analogous to the brane bending that appears in
 T-dual constructions, as we will later review.
 Nevertheless we can still define a parameter by selecting a trajectory via
 \be
\label{diff}
{ \alpha^2_{eff} \over \alpha' }   =  {  M \over 2 }  (t-1) = {   M \over 2 }
\left[  3 \log { \rho  \over  \sqrt{ \alpha' M } } -1  + t_\infty   \right]~.
 \ee
  Thus, we see that once we express $\alpha_{eff}^2$ in terms of the physical size of the two- (or three-) sphere
  at infinity then $t_{\infty}$ appears as an additive constant. Thus we view $t_{\infty}$ as the
  parameter that we can control from infinity.
  In other words, we imagine sitting at a finite but large value of $\rho$ and reading off the value of $t_{\infty}$
  via \nref{diff}. Alternatively we could imagine that we are cutting off the geometry at a large value of
  $\rho$ and embedding it into a compact space.  When we Kaluza-Klein reduce to four dimensions, $t_\infty$ will
  appear as a parameter for the effective field theory describing the small $\rho$ region of the geometry.
  In addition,
  we can view  $\log \rho/\sqrt{\alpha'M}$  as the bare parameter
 at some scale and the left hand side as the scale dependent coupling given by an  RG running.
 At this stage this is not the running of the coupling in
 any decoupled, local, four dimensional field theory. It is a running
 in the four dimensional effective field theory that results from Kaluza-Klein reduction.

 In the regime that $t_{\infty}$ is very large and positive one can show that the solution has a region where
 it looks very close to the resolved conifold with some branes wrapping the $S^2$. As we get close to these
 branes the solution takes into account back reaction and the geometry in this ``near \horizon'' region is
 the Chamseddine-Volkov/Maldacena-Nu{\~n}ez solution \cite{volkovone,volkovtwo,MN}. In this case $\gamma^2 $ is
 very close to one. More explicitly, in the region $t \ll t_{\infty}$ the solution looks like the
 CV-MN solution. In the region $ t \sim t_{\infty}$ the solution looks like the resolved conifold, with some
 branes wrapping the sphere. For larger values of $t$ it looks like the resolved conifold with the
 ``running'' $\alpha^2_{eff}$. The metric is very close to the metric of the resolved conifold for a large
 range of distances when $t_\infty \gg 1$.
  We discuss this in more detail in appendix A.

  On the other hand, when $t_{\infty} $ is large but negative,
   the solution looks like the deformed conifold with
  a very large $S^3$, see figure \ref{regions}(b).
  The size of the $S^3$ is determined by the IR parameter $\gamma^2$, which is becoming very large as
   $t_{\infty} \to -\infty$.
  For this solution the total change in the dilaton is not very large, it is of order $1/\gamma^2$, see
  appendix A. In fact,  for very large $\gamma^2$ it is possible to find an approximate solution
  of \nref{equationc},\nref{equationf} by ignoring the second term in   \nref{equationc}, see appendix A.
   This gives the standard solution for
  the deformed conifold, which in our notation is
\bea
&& \!\!\!\! \!\!\!\! ds^2_{DC} =  { M \alpha' \over 4 } \gamma^2 \Big[ \frac{\sinh^2 t }{K(t)^2} ( dt^2 + (\epsilon_3 + A_3)^2) + { K(t) \over \tanh t} ( \epsilon_1^2 + \epsilon_2^2 + e_1^2 + e_2^2) \notag\\
 &&~~~~~~~~~~~~~~~~~~+ 2 { K(t) \over \sinh t} ( \epsilon_1 e_1 + \epsilon_2 e_2) \Big] \notag\\
&&  ~~~~~{ \rm where }~~~~~
K (t) =  \frac{3^{1/3}}{4^{1/3}} (\sinh 2t - 2t )^{1/3}~.
\label{DCmetric}
\eea

   Thus, this solution displays very explicitly the geometric transition discussed in
   \cite{vafageometric,cachazogeometric}. It is a simple example where the transition happens
   within the supergravity description.
    In fact, for large positive  $t_{\infty}$
     we can view the solutions as branes wrapping the $S^2$ and for large
    negative  $t_\infty$
    we have a deformed conifold with flux, see figure \ref{regions}.
     When the flux is zero, the deformed and resolved conifold represent
    distinct branches. With non-zero flux these branches are smoothly
    connected, see figure \ref{regions} (a). It is interesting
    that one can see this transition purely in supergravity. Of course,
    we knew from \cite{KS} that this transition happened. However, here
    we see it directly in a simpler setting with only one kind of branes and fluxes.
     This system should be contrasted to the closely related problem of D6 branes wrapping the
     $S^3$ of the deformed conifold \cite{AMV,AW}. In that case, the transition could not be seen purely in
     supergravity.

 By comparing the mixed term in the large $t$ expansion of the
metric \nref{themetric} to the same term in the deformed conifold metric \nref{DCmetric}
we can define an ``effective deformation'' parameter
\be
\varepsilon^2_{eff} = \frac{1}{3^{1/3}}e^{-2t_\infty/3}~.
\ee
We see that the effective resolution and deformation parameters are not independent.
 In particular, for very large positive $t_\infty$ we get a large amount of resolution,
 and a small deformation. We depicted this schematically in Figure \ref{regions}(a).
 When $t_\infty$ is very large negative, we get a large deformation parameter.
 However, we cannot use the definition in  \nref{diff}  to measure the effective resolution. In fact,
 at infinity, the effective resolution always becomes big, since it runs.
  Another way to say this, is that in this geometry the \emph{complex} and
(would-be) \emph{K\"ahler} structures are not independent.

 Note that taking $t_{\infty }$ very negative does not correspond to wrapping anti branes on the flopped
 version of the resolved conifold. If we had no flux and we had made $\alpha^2$ negative in \nref{RCmetric}
 this would have been the case. However, in our case, even if we take $t_{\infty}$ very negative we see
 that $\alpha^2_{eff}$ is still growing towards a positive value at infinity.
 The solution where we wrap anti-branes on the flopped version is a separate configuration related by the
 $\Z_2$ operation described above. In some sense, it is connected to the unflopped version by going to
  $t_{\infty } \to -\infty$ (or $\gamma^2 \to \infty $)  and
 back on the other branch of figure \nref{regions} (a).

    When we consider strings propagating on this geometry we have
    (0,2) worldsheet supersymmetry. It
    would be nice to see if there is a gauged  linear sigma model
    which describes this background, in the same way that there is
    one for the conifold with no flux \cite{Witten:1993yc}.

  An interesting object that exists in this geometry is a domain
  wall that comes from wrapping an NS five-brane on the $S^3$. The
  three other directions are an $\R^{1,2}$ subspace of $\R^{1,3}$.
  Its tension is given by
  \be \label{branetension}
   T_2 = { 1 \over ( 2 \pi)^5 \alpha'^3 e^{ 2 \phi(0)} }   V_{S^3 }   =
   { M^{3/2} \over ( 2 \pi)^3 \alpha'^{3/2} e^{ 2 \phi_\infty } } { 1 \over 18 }   e^{-t_\infty  }
  \ee
  where we used that $e^{   2 \phi (0) } = 9 \gamma^3 e^{t_{\infty} } e^{ 2 \phi_\infty} $ \footnote{
  Do not confuse $\phi_0$, which is just an additive constant, with $\phi(0)$ which is the full
  value of $\phi$ at
  $t=0$.}. Note that even though we are evaluating the tension of a brane located at $t=0$, the final
  answer can be expressed in terms of the quantities defined at infinity. This is related to the fact
  that this tension is BPS and that we can compute the superpotential for this configuration in terms
  of the parameters at infinity. As we will see later, this is intimately related to the fact that
  the domain walls tension in the Klebanov-Strassler solution \cite{KS} are independent of the VEV of
  the baryons \cite{DKS,DKT}.

The geometry underlying the  solution can be neatly characterized
in terms of a two-form $J$ and a complex three-form $\Omega$
defining the $SU(3)$-structure, which are constructed from spinor
bilinears\footnote{Explicitly, we have $\Omega_{ijk}= \epsilon^T \Gamma_{ijk}\epsilon$ and $J_{ij}=\epsilon^\dagger \Gamma_{ij}\epsilon$, where $\epsilon$ is the internal spinor solving the supersymmetry equations in string frame.}.
By construction these satisfy the algebraic constraints $\Omega \wedge \bar \Omega = - \tfrac{4i}{3}J^3$
and $J\wedge \Omega =0$.
The conditions imposed by supersymmetry on these forms were derived in \cite{Strominger:1986uh,Hull:1986iu},
and can be written concisely as calibration conditions   \cite{GMW}
\bea
d (e^{-2\phi}\Omega ) & = & 0\label{omegasusy}\\
e^{2\phi}d  (e^{-2\phi}J) & = &    - *_6 H_3\label{jsusy}\\
d (e^{-2\phi}J\wedge J ) & = & 0~.
\label{susy}
\eea
In fact, imposing these equations on a suitable ansatz \cite{PT} is one way to derive the BPS equations
\nref{equationf}, \nref{equationc}.
The  condition \nref{omegasusy} implies that the manifold is complex. In particular we can define a rescaled
three-form $\omehol = e^{-2\phi}\Omega$ which is then a holomorphic (3,0)-form.
The fact that the two-form $J$ is not closed implies that the manifold is not K\"ahler.
For the  solution we are considering we have the following explicit expressions
\bea
 && \!\!\!\!\!\!\!\!\!J =\frac{\alpha' M}{4} \Big[ \left(\coth t (t \coth t - 1) - c  \right) e_1 \wedge e_2 +
\left(\coth t (t \coth t - 1) + c  \right) \epsilon_1 \wedge \epsilon_2\notag\\
&&~~~~~~~~~~~ + \frac{1}{\sinh t } \left( t \coth t - 1\right) \left( \epsilon_1 \wedge e_2 + e_1 \wedge \epsilon_2\right) -  c' dt \wedge  (\epsilon_3 + A_3 )\Big]  \\[2.5mm]
&&\!\!\!\!\!\!\!\!\!\!\!\!\omehol = \frac{\e^{- 2\phi_0}(\alpha' M)^{3/2}}{8} \Big[\sinh t (e_1 \wedge \epsilon_1 + e_2 \wedge \epsilon_2 ) -i \cosh t (\epsilon_1 \wedge e_2 + e_1 \wedge \epsilon_2)  \notag\\
&&~~~~~~~~~~~ - i( e_1 \wedge e_2 + \epsilon_1 \wedge \epsilon_2) \Big]
 \wedge \big( dt  + i (\epsilon_3 + A_3)\big)\label{holothree}
\eea
The holomorphic $(3,0)$-form $\omehol$ is identical to that of the deformed conifold
\cite{Dymarsky:thesis} (see eq. (2.79) of this reference), implying that the solution here has the same complex
 structure as the latter. This agrees with the arguments in \cite{vafageometric} which said that the addition of
 RR fluxes would not change the topological string, which depends on the complex structure, since
 the solution we are discussing is S-dual to a solution with only RR fluxes.

\subsection{The superpotential}

  We can also discuss the superpotential for this solution.
This  is a generalization of the Gukov-Vafa-Witten (GVW)
superpotential \cite{Gukov:1999ya}, and  can be extracted, for example,
from the general expression
in \cite{Grana:2005ny,Koerber:2007xk}, see also \cite{Cardoso:2003af}.
The superpotential is
\be
W~ =~  \int_{M_6} \omehol \wedge ( H_3 + i  d J  )~.
\label{superpot}
\ee
We see that  extremising this superpotential will relate the complex structure $\omehol$ to the would-be
K\"ahler structure $J$, complexified by the $B$-field. We already remarked that this
is indeed the case for our solution.  On the other hand, recall that extremizing
the ordinary GVW superpotential fixes the complex structure of the Calabi-Yau, in terms of the
integer fluxes \cite{Giddings:2001yu}.

Of course, computing the superpotential does not require knowing the
full solution. In fact, it is possible to compute it ``off shell'' by introducing a resolution parameter $S$ and
then extremize it to find the on shell value, as explained  in \cite{vafageometric}.
Here we will simply see how one
recovers the final on shell value from the classical solution. This is a
simple check of the formulas for superpotentials
in the literature and an example that shows how the different quantities
 entering in the definition of the superpotential
look like in an explicit example. The uninterest reader can jump to
the next subsection.

The superpotential \nref{superpot} can be evaluated explicitly
by a  computation as in \cite{cachazogeometric}. Namely, we can use the formula
\bea
\int_{M_6} \omehol \wedge ( H_3 + i  d J  ) ~ = ~ \int_\Gamma
\omehol \cdot \int_\MyUpsilon  ( H_3 + i  d J  )   -  \int_\MyUpsilon \omehol \cdot \int_\Gamma  ( H_3 + i  d J  )
\label{poincform}
\eea
where $S^3$ is the compact three-cycle and $\Gamma \simeq \R^3$ is the dual
non-compact three-cycle. Notice that the  three-forms being integrated are indeed  closed.
We now evaluate the terms in \nref{poincform}. First, let us define two representative
three-cycles as
\bea
&& \MyUpsilon  ~= ~\{ t=0, \theta_1 = \mathrm{constant}, \phi_1 = \mathrm{constant} \}
\notag\\
&&\Gamma  ~= ~ \{ \theta_1 = \theta_2, \phi_1 =  - \phi_2, \psi = \psi(t)\}  ~~~{\rm with}~~ \psi(0)=0~,~~\psi(\infty) =  \psi_\infty
\eea
where $\psi_\infty$ is a constant reference $\psi$. This is an additional parameter of the solution,
which is related to the $\theta$ angle of the associated Yang-Mills theory via
$\theta \sim M \psi_\infty $. The reason it is a parameter is because the integrals will depend on
$\psi_\infty$. We then have that
\be
\Omega|_{S^3} = e^{ 2 \phi} \omehol |_\MyUpsilon ~ =~ e^{ 2(\phi(0) -\phi_0) }
  { ( M \alpha')^{3/2} \over 8 } \epsilon_1 \wedge \epsilon_2 \wedge \epsilon_3 =  dv_3
\ee
where $dv_3$ is the volume element of the cycle in string frame.
This shows that the three-sphere $S^3$ is calibrated, i.e. it is a supersymmetric cycle.
The other fluxes are
\bea
\int_\MyUpsilon ( H_3 + i  d J  ) = - 4 \pi^2 \alpha' M ~,\quad
\int_\Gamma ( H_3 + i  d J  )  = \int_{S^2_c} (B+i J) \equiv b + ij
\eea
where $S^2_c$ is a two-sphere at some cut-off distance $t_c$. The periods of the holomorphic three form are
\bea
\int_\MyUpsilon \omehol &=&
\frac{(\alpha' M)^{3/2}}{8}  \frac{16\pi^2}{9e^{2\phi_\infty}}e^{-t_\infty - i \psi_\infty}~, ~~
\\
\int_\Gamma  \omehol &=&
 \frac{(\alpha' M)^{3/2}}{8} \frac{4\pi i}{9e^{2\phi_\infty}}e^{-t_\infty - i \psi_\infty} \left[
 2(t_c + i \psi_\infty)   + e^{-t_c - i \psi_\infty}- e^{t_c + i \psi_\infty } \right]
\label{periods}
\eea
where we have multiplied the expression for $\omehol $ in \nref{holothree} by $e^{-i \psi_\infty}$ to make
sure that it depends holomorphically on the parameter $t_\infty + i \psi_\infty$. We are free to define
the phase of the three form.
The period on the non-compact cycle contains a divergent term, however after changing  variables
as
\be
e^{ t - t_\infty}  = \frac{\rho^3}{(\alpha' M )^{3/2}}
\ee
we see this term does not depend on the parameter $t_\infty$, thus it can be dropped, as in
\cite{cachazogeometric}.  As we discussed around \nref{diff} we can define a bare scale parameter
as  $\hat \Lambda = \rho_c /\sqrt{\alpha' M}$, and then, to compare directly with
with \cite{cachazogeometric},  let us also define $S=e^{-t_\infty - i \psi_\infty }$
and a ``complex coupling'' $2\pi \alpha'  \tilde \alpha = -(j+i b)$.
Then we have
\bea
W = (\alpha' M)^{3/2}  \frac{4\pi^3 i\alpha'}{9e^{2\phi_\infty}}
\left[ -  M S \log {\hat \Lambda^3\over S} - S \tilde \alpha\right]~.
\label{vylike}
\eea
The prefactor can be absorbed in the definition of $S$.
This has the expected form of the Veneziano-Yankielowicz
 superpotential, if we regard $\tilde \alpha$ as the running coupling as in
\cite{cachazogeometric}. Indeed,
in the field theory limit we discuss below
the real part of $\tilde \alpha$ may be interpreted
as the 4d gauge coupling of the ${\cal N}=1$ SYM theory, and $S$ is identified with the glueball superfield
\cite{vafageometric}.
Here we can evaluate $\tilde \alpha$ in the solution at hand, getting
\be
\tilde \alpha = M( t_c -1 +i \psi_\infty ) =
M( t_\infty + \log \hat \Lambda^3  -1  + i \psi_\infty )
\ee
where we have included a contribution from a flat $B$-field at infinity.
Now, inserting this into \nref{vylike}, the logarithmically divergent terms correctly cancel and
 we get
\be
W = (\alpha' M)^{3/2}\frac{4\pi^3 i}{9e^{2\phi_\infty}}M\alpha' e^{-t_\infty - i \psi_\infty } ~.
\label{weval}
\ee
This agrees\footnote{We have to divide by  $\alpha'^4$  in order to restore the correct units in the superpotential.}
with the domain wall tension \nref{branetension},   noting that for large $M$,
$T_2 \sim |\Delta W|\sim W/M$.

The solution depends on the parameter $\Phi = M ( t_\infty + i \psi_\infty)$ which has the identification
$\Phi \sim \Phi + 2 \pi i $. The fact that it is not invariant under the naive shift symmetry of $\psi$ is
related to the $U(1)_R$ breaking as discussed in more detail in \cite{MN}.
The pattern of breaking of this  $U(1)_R$  is the simplest way  to derive \nref{weval}
\cite{Intriligator:1995au}.

 Now let us discuss a particular limit of this configuration which is supposed to lead to a decoupled
 ${\cal N}=1$ pure Yang-Mills theory.
When $t_\infty $ is very large and positive,
  it can be interpreted as the size of the $S^2$ that
 the NS five branes are wrapping, see appendix A.
   In that case the full solution looks as in figure \ref{regions}(c).
 Notice that if we reduce an NS fivebrane on an $S^2$ of radius $\alpha$, which is very large,
 then  the four dimensional  gauge coupling\footnote{
  In this limit,  the above superpotential is the one arising from gluino condensation
  $W \propto \Lambda^3 $.  In particular, one gets the correct coefficient for the beta function,
 see also \cite{DiVecchia:2004dg}.} is given by
 \be  \label{uvcoupling}
 {8 \pi^2 \over g_4^2 } = 2  { \alpha^2 \over \alpha' } =M t_{\infty}
 \ee
 where we have evaluated the radius of the two-sphere at the value that it has in the region where the
 solution looks similar to that of  the  resolved conifold. This is possible only if $t_{\infty}$ is very large.
 This is the value of the coupling at the Kaluza-Klein scale set by the radius of the sphere. Note that
 $t_{\infty}$ gives the 't Hooft coupling and it parameterizes the gravity solution, as expected.
 This coupling has to be very small, thus requiring that we take $t_{\infty} \to \infty$.
 In addition, we would like to decouple the fundamental strings. A fundamental string stretched along one
 of the non-compact four dimensional directions is a BPS state in this geometry. It can sit at any value
 of the radial coordinate. We want these strings to be much heavier than the branes discussed in
 \nref{branetension}. This can be achieved if we keep $\phi_\infty$ fixed as we take $t_\infty \to \infty$.
 At the origin we find that $\phi(0) \to \infty $ in this case. Thus, as
  expected, we should S-dualize to the D5 brane picture in order to analyze the limit.
   The limit we are taking is such that the S-dual coupling is becoming extremely small and that the same time
   the size of the $S^2$ is also becoming small in the new string units, but with \nref{uvcoupling} still large.
   This decouples the D1 branes which are S-dual to the BPS fundamental strings mentioned above.
   Of course, in this regime the gravity solution fails and would probably have to use non-critical strings
   to describe the large $M$ limit of the theory.
    Nevertheless, the superpotential computed in terms of $t_{\infty}$ continues to be valid since it is
    independent of $\phi_\infty$. So we can first take $\phi_\infty$ small enough so that $\phi(0)$ is not so large
    and we can trust the gravity description. Then we take $\phi_\infty $ to a larger value so that the field theory decouples.
    Now this discussion seems to be explicitly contradicted by the fact that the tension \nref{branetension} depends
    on $\phi_\infty$. However, this dependence is simply a choice of units. In fact, it could also be viewed as arising
    from the K\"ahler potential in a situation where we compactify the theory and go down to four dimensions. Since
    those terms in the K\"ahler potential are determined in the bulk region of the six dimensional space, they are not
    corrected by the physics in the tip.

Note that the system of $M$ branes on $S^2$  naively
has an overall $U(1)$ gauge symmetry. This mode becomes
non-normalizable in the solution. If we were to start from our configuration with 5 branes on $S^2$
 then we
can add the flux of this $U(1)$ gauge field along the four dimensional space by
performing a U-duality. In this case we see that the asymptotic form of the metric
changes in a non-normalizable way.  Namely,
a $U(1)$ flux on the branes   induces lower brane charges which  contribute
to the logarithmic running of the resolution  parameter.

\subsection{Relation to brane constructions}

\label{branepict}

   The conifold is T-dual to two orthogonal NS branes.
   In other words, we have an NS brane along 012345 and an NS' brane along 012367.
   Strictly speaking we should introduce a compact direction along which to do the T-duality, see
   \cite{Dasgupta:1998su} for further discussion.
   The compact direction can be the direction 8. Thus we can have these branes separated along the
   direction 9. We have a cylinder formed by directions 8 and 9.
   This configuration has one more parameter, relative to the conifold, which is the radius in the
   8 direction. In the limit that the radius, $R_8$,  in the 8 direction goes to zero we expect to recover the
   conifold after a T-duality\footnote{It would be nice to derive the geometry that is T-dual to the
   system of fivebranes for a finite value of $R_8$. }.
    Thus the brane picture contains yet one more parameter.
   We have the string coupling, the radius of the 8th direction and the separation between the branes.
    We then consider a sort of near \horizon limit of the fivebranes where we take $r\to 0$ and $g_{IIA} \to 0$
    with $ g_{IIA}/r$ fixed. At the same time we take $R_8$ to zero so that
    $g_{IIA} { l_s \over R_8}  = g_{IIB}  $  is kept fixed. Here $g_{IIB}$ is the value of the IIB coupling.

\begin{figure}[ht!]
\epsfxsize = 9cm
\centerline{\epsfbox{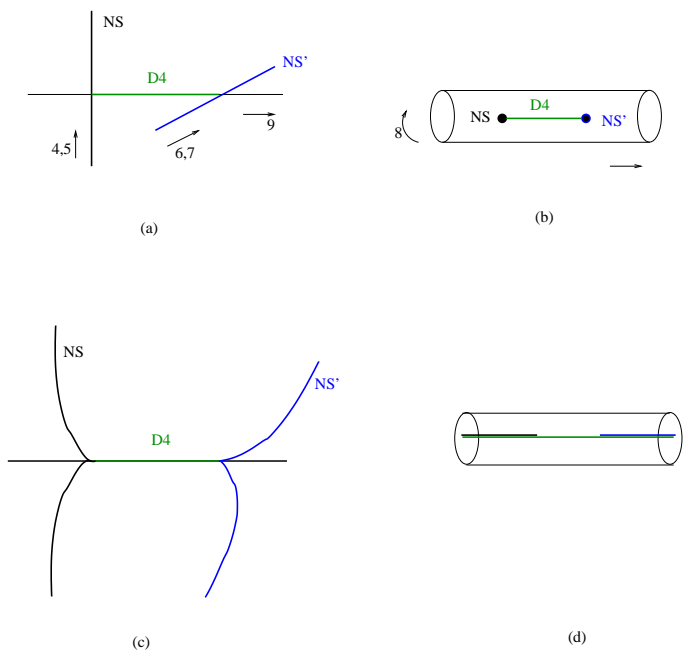}}
\caption{(a) A D4 brane stretched between two orthogonal NS fivebranes. In (b) we compactify a direction
orthogonal to all the branes in (a). In the limit that the size of the 8th circle goes to zero we expect
to recover the conifold. In (c) and (d) we have schematically represented the effects of brane bending.
The transverse position of branes varies logarithmically. This has the same origin as the dependence
of the parameter $\alpha_{eff}^2$ on the radial position.}
\label{branepicture}
\end{figure}

   When we add D4 branes stretching between the NS branes we find that the NS branes bend and there
   is a logarithmic running of the separation between the branes, see figure \ref{branepicture}.  See also
   \cite{Elitzur:1997fh}.

    Another possibility  is the M-theory construction in \cite{Witten:1998jd,Hori:1997ab}.
    That is obtained in the limit that
    $R_8 \to \infty$ and $g_{IIA} \to \infty$. Thus, adding the size of the extra circle $R_8$ as a parameter
    allows us to interpolate between the various pictures that have been proposed for describing ${\cal N} =1$
    SYM.

    In particular, the superpotential of the theory, as a holomorphic function of the non-trivial parameter in
    all these pictures is expected to be the same because we do not expect any dependence of the superpotential
    on the parameters that we are varying. This is due to the fact that the partners of the parameters that
    we are varying are axions and we do not have any finite action BPS
    instantons which could contribute to the superpotential.

Let us discuss the various moduli in the type IIB picture, from the point of view of
the geometric side. This would be a geometry similar to the conifold,
except that it asymptotes to $S^1 \times \R^5 $ at infinity, since the size of the 8th direction is finite.
 We are considering the background
in which we wrap D5 branes, S-dual to the original solution. We can consider how 
various fields are paired under the supersymmetry preserved by the D5 brane wrapped on the 
compact two cycle. 
The string coupling, $g_s$, is paired by supersymmetry to the RR axion, $a^{RR}$, dual to
$C_2^{RR}$ in four dimensions, i.e. $*_4 d a^{RR}= F_3^{RR}$.
The corresponding instanton is a D5  fivebrane wrapped over
all the six internal dimensions.
The radius of the 8 direction $R_8$ is paired to a RR $C_2$ field along the internal directions.
 The corresponding instantons are non-compact euclidean D1  branes  extended along the
internal directions and wrapped along  the eighth dimension at infinity.
The ten dimensional RR axion is paired with the
four dimensional NS axion $a^{NS}$ which is dual to $H_3^{NS}$ with all four dimensional indices.
The corresponding instantons are NS5 branes along the internal dimensions.
  Finally, we have $B^{NS}$ on the compact two cycle which is paired with $C_4$ on a  non-compact
internal four-cycle. The corresponding euclidean D3 brane instantons also have infinite action.
Thus, holomorphy, plus the absence of finite action instantons, imply that  the superpotential is only a function of
$t_\infty + i \psi_\infty$. Thus, we can vary the other variables from  the values which decouple the four dimensional
theory to other values where the M theory brane picture is a good approximation.

    In the case in figure \ref{branepicture} (b) we have a BPS string corresponding to a D2 brane wrapping the
    8th direction, see figure \ref{bpsstrings}(a).
     This T-dualizes to a D1 brane on the original picture, which is a  state that we
    have to decouple to get to the ${\cal N}=1$ pure Yang-Mills theory.

\section{Solutions with D3 branes from a duality transformation}

In this section we recover the  solution in \cite{Butti}
by applying a simple chain of dualities
to the solution  discussed in section \ref{section_one}. This introduces various fluxes.
In fact, the procedure that we discuss
is quite general, and it can be applied to any solution  with
only dilaton and NS three-form turned on. In principle, the starting solution may also
be non-supersymmetric. However, if it preserves supersymmetry and is therefore
of the type discussed in \cite{Strominger:1986uh}, then the duality maps it to
a supersymmetric solution of type IIB with non-trivial  NS and RR
fluxes, where the internal six-dimensional geometry is of $SU(3)$-structure
type\footnote{The duality
may be easily adapted  to other supersymmetric geometries of the type  $\R^{1,d}\times M_{9-d}$
with dilaton and NS three-form \cite{GMW},
producing corresponding supersymmetric solutions with non-trivial RR fluxes, both in type IIA and type IIB.}.
The internal geometry is not (conformally) Calabi-Yau, and the three-form fluxes are not
imaginary self-dual.
However, for the solution that we discuss here, we will see that the latter
may be recovered by taking a certain limit.  In particular, in this limit we  recover the Klebanov-Strassler
warped deformed conifold geometry \cite{KS}.

Let us now describe the dualities. First of all, we perform an S-duality on the initial solution,
which then represents D5 branes wrapped on the $S^2$ of the resolved conifold.
 The solution has non-trivial dilaton  and a RR three-form flux. We then compactify
 on a torus three spatial world-volume coordinates of the D5 branes
and perform T-dualities along these directions, obtaining a type IIA configuration of
D2 branes wrapped on the $S^2$. This is then uplifted to M-theory, where we do a boost\footnote{The coordinate $t$ here is the time coordinate, and should not be confused with the variable $t$ elsewhere in the paper.}
\bea
t  \to
~\;\,\cosh \beta \,t - \sinh \beta \,x_{11}~, \qquad x_{11}  \to  - \sinh\beta \,t + \cosh\beta \,x_{11}~,
\eea
obtaining a configuration with M2, and Kaluza-Klein momentum charges.
 Finally, we reduce back to type IIA and repeat the three T-dualities on the torus.
The resulting type IIB solution has  D5-brane plus  D3-brane charges\footnote{Of course, the D3 Page charge is still zero.
See \cite{Benini:2007gx} for further discussion of various definitions of charge in this background. }.
The steps involved in the transformation are summarized
by the following diagram
\bea
D5 ~\to ~D2 ~\to ~M2 ~\to~ M2,  p_{KK} ~\to D2,  D0 ~\to~  D5, D3
\eea

The final result is the following solution
\bea \label{f5solution}
\hat \phi = \phi_{here} &=& - \phi_{\rm previous } = - \phi \\
 d\tilde s^2_{str}
&=& \frac{1}{h^{1/2}}dx_{3+1}^2 + { e^{ \hat \phi_\infty} \tilde M \alpha' \over 4 } { h^{1/2} \over \cosh \beta }  e^{-2(\phi - \phi_{\infty})}ds^2_6
\\
h & = & 1 + \cosh^2 \beta (e^{2(\phi-\phi_\infty)}-1)
\\
F_3 & = &   {\alpha' \tilde M \over 4} w_3 \qquad
H_3 ~ = ~ -
\tanh \beta  {  e^{  \hat \phi_\infty} \tilde M \alpha' \over 4 } e^{-2(\phi - \phi_{\infty})}
*_6 w _3  \label{f3rr}
\\
F_5 & = & -\tanh\beta e^{- \hat \phi_{\infty}} (1+*_{10})\vol_4\wedge
dh^{-1}
\eea
The six dimensional  metric $ds^2_6$  and the three form $w_3$ are  the same as in
\eqref{ansatz}. Notice that the dilaton here is minus the dilaton in  \eqref{ansatz}. We denote
by $\hat \phi$ the dilaton for this solution in this frame and we continue to denote by
$\phi$ the expression for the dilaton in \nref{ansatz}.
Note that all the terms involving $\phi - \phi_\infty$ do
not depend on the constant $\phi_0$ in \nref{ansatz}. So we should think of $\hat \phi_\infty$ as
a new parameter determining the asymptotic value of the coupling. Here $F_3$ denotes the
RR three-form and $H_3$ the transformed NS three-form.
 The parameter $\tilde M$ is the quantized RR flux through a three-sphere at infinity,
  representing the number of D5 branes that we are wrapping. This is related to the parameter $M$,
  giving the number of NS fivebranes in the original solution as
\bea
 \tilde M ~=~ \frac{1}{4\pi^2\alpha'} \int_{S^3_\infty} F_3 ~= ~  e^{ \hat \phi_{\infty} }
 M \cosh\beta \in \mathbb{N}~.
\eea
Thus in particular the original gravity parameter  $M$ is not quantized in the transformed solution\footnote{
This is a usual feature of supergravity duality transformations which are a symmetry of the gravity equations but not
a symmetry of the full string theory. }.

We see that the changes of the solution with respect to the one in
section \ref{section_one} are rather simple. However, crucially
non-trivial RR fluxes are generated.  Note that the three-form
fluxes satisfy the relation
\bea
 \cosh \beta H_3 + \sinh \beta *_6
e^{-2(\phi - \phi_{\infty} )} e^{\hat \phi_{\infty} }  F_3 \,= \, 0
\label{genisd}
 \eea
which is a
generalization of the imaginary-self-dual condition for
supersymmetric fluxes on a Calabi-Yau geometry \cite{Giddings:2001yu}.
The boost
parameter $\beta$ can be thought of as an interpolating parameter
\cite{Frey:2003sd}. When this goes to zero, the solution reduces
back to the initial one. On the other hand, as we discuss below,
in a certain limit of infinite boost, one can recover the warped Calabi-Yau
solution, with  imaginary self dual fluxes.  Notice that the NS $H_3$ is
manifestly closed. This follows from the fact that in the initial
NS5 solution, this obeys the calibration condition \nref{jsusy}.
Then  we can read off the $B$ field in terms of the two-form $J$,
namely
\bea
B \, = \, \sinh \beta e^{-2\phi} J~.
\label{Bfield}
\eea

The warp factor $h^{-1/2}$ is an increasing
function of $t$ which goes to one at infinity.
{}From the small $t$ expansions of the functions given in section \ref{section_one},  we see that
the warp factor $h^{-1/2} $ becomes constant at $t=0$, thus the IR geometry is essentially the same for
all values of the parameter $\beta$.
In particular, there is a finite-size $S^3$.
D5-branes wrapping this $S^3$  corresponds to a domain wall in the four-dimensional world-volume.
It can be shown that this is a BPS object, because the $S^3$ is calibrated by the generalized special Lagrangian
calibration, namely the three-form $\Omega$ discussed in section \ref{section_one}.
The tension of this domain wall is in fact equal to \nref{branetension}.
However, after expressing it in terms
of the physical $\tilde M$, we have
\bea
T_{DW} ~= ~ \frac{\tilde M^{3/2} e^{\hat \phi_\infty/2}} {(2\pi)^3 \alpha'^{3/2}}
 \frac{1}{18   \left(   e^{ { 2 \over 3 } t_\infty } \cosh \beta  \right) ^{3/2}}~.
\label{dwtension}
\eea
Notice that this depends on the parameters $\beta$ and  $t_\infty$,
but only through a particular combination. We will further comment
on this dependence below.

The solution \eqref{f5solution} is contained in
\cite{Butti}. To obtain the baryonic branch solution they have set an integration
constant to a
particular value\footnote{We have that the constant $\eta$ in \cite{Butti} is $\eta= - \tanh \beta $.}. Here
we have restored it to its more general value.
In this more general solution, the warp factor goes to constant at infinity.
Thus we are coupling the field theory to the string theory modes of the ordinary conifold.
In particular, we are also
gauging the baryonic $U(1)_B$. Then, a combination of the parameters $t_\infty$ and $\beta$ may be thought
of as the value of the FI parameter for this $U(1)$,
while another combination corresponds to the domain wall tension \nref{dwtension}.
The baryonic branch interpretation can be recovered in a limit in which we send $\beta$ to infinity.
We will  discuss momentarily the relationship of the domain wall tension above
 and that computed in \cite{DKS,Dymarsky:thesis,DKT}.

Again, we can discuss the generalized GVW superpotential for this solution.
For the type of geometry we have  the general expressions in \cite{Koerber:2007xk,Grana:2005ny}
reduce to the  simple form\footnote{The geometry we have is of the simpler $SU(3)$ structure type,
as opposed to the more general $SU(3)\times SU(3)$ structure type. Here we are setting the type IIB RR axion to zero.}
\be
W~ =~  \int_{M_6} \omehol \wedge (F_3 + i e^{-\hat \phi} \cos w H_3  + i  d(\sin w J)  )~,
\label{gensuperpot}
\ee
where we expressed it in terms of the same forms $\omehol$ and $J$ of the previous section. The function
$w$ is the same appearing in \cite{Butti}. This  may be thought of as
 part of the data defining the $SU(3)$-structure, in addition to $J$ and $\Omega$.  In particular, it is a degree of freedom in the spinor ansatz that solves the type IIB supersymmetry equations\footnote{In the formalism of generalized geometry adopted in \cite{Koerber:2007xk},  $e^{iw}$ is the zero-form part of the pure spinor $\Psi_1$.}.  We have that
$\cos w = - \tanh \beta e^{\hat \phi- \hat \phi_\infty}$, thus it is
suggestive to re-express the boost parameter in terms of
an angular parameter, defining $\sin \delta = - \tanh \beta$, so that
\be
\cos w = e^{\hat \phi- \hat \phi_\infty} \sin \delta~, \qquad \sin w = e^{ \hat \phi- \hat \phi_\infty } h^{1/2}\cos \delta
\ee
and the superpotential becomes
\be
W~ =~  \int_{M_6} \omehol \wedge \left[ F_3 + i e^{- \hat \phi_\infty} \left(\sin \delta H_3  +   \cos \delta
d(e^{\hat \phi}h^{1/2} J) \right) \right]~.
\label{gensuperpot2}
\ee
We see that this form interpolates between the original GVW superpotential when $\cos \delta = 0$ and
the S-dual version of the one discussed in the previous section, when $\sin \delta = 0$. Although the limit
$\sin \delta \to 0$ is straightforward, the infinite boost limit $\cos \delta \to 0$ should be done more carefully, but it does reproduce the correct GVW expression in the KS solution.
We could also write  the superpotential \nref{gensuperpot2} using the formula \nref{poincform} as before.
Then the discussion is essentially unchanged, provided we replace $b + i j \to c_2 + i (e^{-\hat \phi_\infty}\sin \delta b + \cos \delta j )$ and write the periods of the holomorphic three-form in terms of $\tilde M$.
In the end the superpotential depends only on the parameter
\be \label{defofl}
L = { U\over \cos \delta} = 12 e^{2/3t_\infty}\cosh \beta
\ee
and we recover the domain wall tension \nref{dwtension}.
  Notice that the effective ``coupling constant'' is then
\bea
\frac{8\pi^2}{g_4} = \frac{1}{2\pi \alpha'}(\cos \delta \, \,  j + \sin \delta \, \, \frac{b}{g_s})~.
\eea
This interpolates between the definition we discussed in section \ref{section_one} for $\delta =0$, and the definition of the coupling
$g_-^2$ used in \cite{KW,KS} for $\delta = \pi/2$.

In fact it is rather natural to change the angle $\delta$ keeping $L$ fixed. In the limit
that $\cos \delta \to 0$ we obtain a finite limit which is simply the Klebanov-Strassler solution
but with a +1 in the warp factor, so that the warp factor asymptotes to one at infinity. The reason
we get a smooth solution in this limit is due to the fact that in this limit the function
$e^{ 2 \phi - 2 \phi_\infty} -1 \sim U^2 \times $(finite), as shown in appendix A.

\subsection{Recovering the Klebanov-Strassler asymptotics}

The solutions we discussed so far are such that the warp factor goes to one at infinity, which is a reasonable
thing to consider.

However, one can consider a near \horizon limit  where the warp factor grows without bound as we go to large $t$.
This gives the  solution \cite{Butti}
 with Klebanov-Strassler \cite{KS} (or Klebanov-Tseytlin \cite{Klebanov:2000nc}) asymptotics.
This can be obtained from the metric above by taking $\beta \to \infty$.
  To obtain a finite
limit we also rescale the worldvolume coordinates
\be \label{rescalex}
x \to e^{ \hat \phi_\infty \over 2 } \sqrt{\tilde M \alpha' } \sqrt{U}  \sqrt{ \cosh \beta } \Lambda_0  x
\ee
where we have included additional factors. The factor of $U$ will make sure that the asymptotic
 form of the metric is independent of $U$. The factor $\Lambda_0$ simply introduces a scale, which is the
 scale of the last step of the cascade.  The other factors are just for convenience.

This then gives the solution
 \bea
 \hat \phi &=& - \phi   \\ \label{KSasymp}
 d\tilde s^2_{str}
&=& e^{ \hat \phi_\infty} \tilde M \alpha'
\left[ \frac{1}{\hat h^{1/2}} U \Lambda_0^2 dx_{3+1}^2 + {  1  \over 4} \hat h^{1/2} e^{-2(\phi -
\phi_{\infty})}ds^2_6 \right] \label{KSmetricfi}
\\
\hat h  & = &  e^{ 2(\phi- \phi_\infty ) } -1
\\
F_3 & = &    {\alpha' \tilde M \over 4} w_3 \qquad H_3 ~ = ~ -
  {\alpha' \tilde M   e^{\hat \phi_\infty}  \over 4} e^{- 2(\phi - \phi_{\infty})}  *_6 w_3\label{f3rrsec}
\\
F_5 & = & -  \Lambda_0^4 ({ \tilde M }\alpha')^2  U^2 e^{ \hat \phi_{\infty}} (1+*_{10})\vol_4\wedge
d \hat h ^{-1}
\eea
where $\phi$, $w_3$ and $ds_6^2$ are the same as in \nref{ansatz}.
This way to write the metric shows clearly the dependence on $\tilde M$ and $\hat \phi_\infty$. They
 appear as simple overall factors.  This implies
 that any gravity computation gives an
answer which scales like $\tilde M^4 e^{ 2 \hat \phi_\infty} = \tilde M^2 ( e^{\hat \phi_\infty} \tilde M)^2 $,
 since this is the  overall factor in the
action. The fact that the metric has a $U$ independent asymptotics comes from the fact that
$\hat h   \propto U^2 t e^{ -4 t \over 3 }$, and $c \propto U^{-1} e^{ 2 t\over 3 } $
for large $t$, see \nref{dilsub} in appendix A. This implies that the dependence on $U$ cancels for large $t$.
Of course the full solution depends on $U$. In fact, this rescaling introduces a factor of $U$
in the term corresponding to the running deformation parameter
\be
\left( { t \over \tanh t } -1 \right) \to U  \left( { t \over \tanh t } -1 \right)~.
\ee
This deformation, which was not normalizable when the warp factor was constant asymptotically, is
now normalizable. In fact $U$ is parametrizing the VEV of the scalar operator that is an
${\cal N}=1$ partner of the baryonic current \cite{DKS}.
The rescaling \nref{rescalex} also has the effect of making the domain wall tension independent of $U$
\be \label{tensionKS}
T_{DW} \propto   { \tilde M^3 e^{ 2 \hat \phi_\infty} \Lambda_0^3 } \sim \Lambda^3~.
\ee
This result was obtained numerically in
\cite{DKS} and then analytically in \cite{DKT,Dymarsky:thesis}. It is simply the statement
that the superpotential is constant along the baryonic branch. Here $\Lambda^3$ is the usual
holomorphic $\Lambda$ which is introduced for this theory \cite{DKS}.

The metric is closely related to the one for the simpler case with only fivebrane charge.
In particular,
  the fact  that the large $U$ region is related to branes wrapped on the resolved conifold
continues to be valid, but with some modifications. The boosting that we have done induces a large, but finite
amount of D3 brane charge on the fivebranes which is proportional to $t_\infty$ as we will show below.
 (Of course, the fluxes lead to a diverging
D3 brane charge at infinity.)  Thus, the theory on the fivebrane becomes non-commutative \cite{SWnoncomm}.
Nevertheless, for large values of $U$, this description in terms of D5 branes wrapped a resolved conifold becomes better
and better.

In the next section we show that this picture also emerges from the field theory analysis in \cite{DKS}, the baryonic
VEVs give rise to a fuzzy sphere which is building up the fivebrane wrapping the $S^2$.
We will discuss there a more detailed comparison with the gravity description. The geometry in the large $U$
region is divided into two parts, one region looks like the resolved conifold, namely the solution of Pando-Zayas and Tseytlin \cite{PZT}.
Near the origin of the resolved conifold, one has a region that looks like the near horizon geometry
of fivebranes. More details are given in appendix A.

\subsection{Parameters of the solutions}

It is interesting to discuss a bit more explicitly the parameters of the various solutions.
For the solutions of section \ref{section_one}, before we perform the boost, we had one discrete parameter $M$
labeling the number of branes and two continuous parameters $\phi_{\infty}$ and $t_{\infty}$. The
solution depended non-trivially only on $t_{\infty}$.

For the boosted solutions we now continue to have a discrete parameter $\tilde M$ which is the net
number of fractional branes. We have a   simple parameter $\phi_\infty$ and
a non-trivial parameter $t_{\infty}$ plus a parameter $\beta$.
These are the parameters in the case that the warp factors asymptotes to a constant. All of these
parameters are non-normalizable.   Notice however that the superpotential and domain
wall tension depend only on the combination $L$, which is paired by supersymmetry
 with the phase coming from the RR $C_2$.

If we further take the limit that leads to Klebanov-Strassler  asymptotics, we loose the parameter $\beta$. Furthermore,
in that case, the parameter $U$ is normalizable and is interpreted as the baryonic branch VEV.
It is interesting to understand how the parameter $t_{\infty}$ becomes non-normalizable once we  change
the asymptotics of the solution. The KS solution also has a $U(1)_B$ global symmetry which
is spontaneously broken. When we change the asymptotics from Klebanov-Strassler to a constant warp factor
at infinity, we are gauging this $U(1)$ symmetry and adding an FI term for this $U(1)$. Thus, when we set
the D-term to zero we relate $U $ to the FI term for this $U(1)$ symmetry. We still have a parameter
that we can change, which is the FI term, which in turn changes $U$, but it is now a non-normalizable parameter.

 Notice also that in the solution  with KS asymptotics,
 the superpotential and the tension of the domain wall do
not depend on $U$. Indeed, with these asymptotics the susy partner of $U$ is the zero-mode of
the RR potential $C_4$ \cite{Gubser:2004qj}, which manifestly does not appear in the superpotential.

\subsection{Brane picture}

\begin{figure}[ht!]
\epsfxsize = 9cm
\centerline{\epsfbox{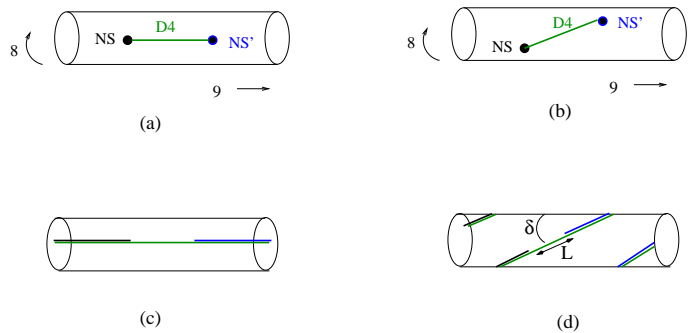}}
\caption{(a)  IIA configuration with a D4 brane stretching between two orthogonal NS fivebranes.
 (b) We separate the fivebranes in the  compact 8th dimension. In (c) we add the effects of brane bending,
 the NS fivebranes
 bend in the non-compact 9th direction. (d) When we include the effects of brane bending the branes now stretch along a slanted
 direction parametrized by an angle $\delta$.
   }
\label{braneslanted}
\end{figure}

We can also  understand the boosting procedure in the type IIA brane picture discussed in subsection \ref{branepict}.
Recall that the solution with only  D5 brane charge   corresponds to the following IIA picture:
 an  NS fivebrane
along directions 012345 and one NS' fivebrane along 012367, with M D4-branes stretching on a segment
 along the direction 9. The remaining direction, 8, is a compact direction. We can now imagine
moving  the NS' brane along the 8th direction, keeping  fixed the NS brane.
Before taking into account backreaction, we have the picture in figure \nref{braneslanted}(b). Here we naively
have two parameters, the separation in the 9th direction and the separation in the 8th direction.
However, once we take into account the effects of brane bending, the parameters really become a choice of RG trajectory,
which can be viewed as the parameter $L$ and an angle $\delta$. We can think of $L$ as a kind of renormalized distance along
the direction where the branes are separated.  In this picture
we see that $U = L \cos \delta$ is simply the projection of the parameter $L$ on to the 9th direction, after we identify
$L$ as in \nref{defofl}.
 Once we have a non-zero angle, the D4 branes   wrap the 8th direction an infinite amount of times, due to the brane
 bending. This translates into the ever growing D3 charge we have in the IIB geometry. Note that, even in the case
 that the warp factor goes to a constant, we have this ever growing D3 brane charge.

Supersymmetry continues to pair the renormalized parameter $L$ with the flux of the $C_2^{RR}$ field on the $S^2$ in the
IIB picture.

The solution with KS asymptotics, on the other hand, corresponds to  rotating by 90 degrees, keeping the distance $U$ fixed.
Notice that the  superpotential depends only on the distance between the NS branes, and not on the amount of rotation.
The configuration of two NS branes displaced along the compact direction is known to corresponds to the
KS solution \cite{KS,Dasgupta:1998su}.
 The
 boosted solution corresponds to a general rotated configuration of NS-NS'-D4.
 In fact, the existence of such solution was anticipated in \cite{Polchinski:2000mx}.

\begin{figure}[ht!]
\epsfxsize = 9cm
\centerline{\epsfbox{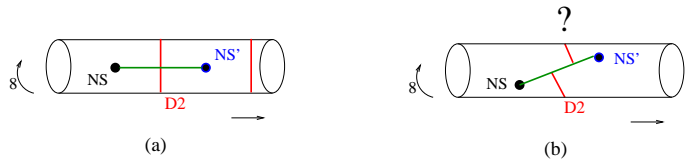}}
\caption{(a)     BPS state which corresponds to a D2 brane wrapping the
 circle. (b) This brane picture would mislead us to expect also a BPS state for a D2 wrapping the circle. However, there is
 no such BPS state in the full theory.
   }
\label{bpsstrings}
\end{figure}

Finally, let us comment on the fate of the BPS four dimensional string that we had for zero angle. Naively, one would
expect that for a finite angle one should continue to have this BPS state. Indeed, the brane picture suggests
a very natural candidate as shown in figure \nref{bpsstrings}(b). This would correspond to a string closely related to the D1 brane in these geometries \nref{f5solution}.
However, the analysis in \cite{Gubser:2004qj} showed that the
D1 brane is not BPS, and no other BPS strings
were found\footnote{Note, however, that in the case that $\tilde M=0$, so that
we study the Klebanov-Witten theory \cite{KW} in its baryonic branch
\cite{baryonicKW}, then such BPS strings do exist \cite{stringsKW}.}.
Naively, from the field theory point of view, one would expect to find such
BPS strings in the case that we gauge $U(1)_B$, which
is what is happening when we have an asymptotically constant warp factor.  In fact, the classical theory
contains such strings \cite{TongVortex}. However, due to the quantum deformation of the moduli space these strings cease to
be BPS in the quantum theory, see section 4.2 of
\cite{TongVortex}. There  strings are D1 branes in \nref{f5solution}, which are not BPS if $\delta \not =0 $.


\section{The baryonic branch and the fuzzy sphere}

We have seen that the large $U$ asymptotic form of the Butti et al solution \cite{Butti}, with Klebanov-Strassler
boundary conditions, can be represented accurately in terms of $M$ fivebranes wrapping the resolved conifold
with a large amount of dissolved D3 brane charge. This gives a good picture for the asymptotic form of
the solution far along the baryonic branch. In this section we drop the tilde in $\tilde M$ and denote it simply by $M$.

In this section we will see that a  field theory analysis of the baryonic branch also leads to
this picture. The field theory analysis of the various branches of the KS theory was done in
detail by Dymarksy, Klebanov and Seiberg in \cite{DKS}. In particular, these authors found
the vacuum expectation values of the fields  along the baryonic branch. In this section we   argue
that these VEVs represent a fuzzy two-sphere. This two-sphere is building up the D5 brane.
In fact, a closely related discussion was also developed  in \cite{Nastase:2009ny} for vacua in the ABJM theory
\cite{abjm}.
This is not a coincidence since the ABJM theory is closely related to the Klebanov-Witten \cite{KW} field theory.

We  start by writing explicitly the
  classical solutions  \cite{DKS} for  the baryonic branch in the weakly coupled version of the quiver field theory.
  We consider the quiver field theory with gauge group $SU( M k ) \times SU(M(k+1) )$.
  For the time being $k$ is fixed and we will discuss later the effects of the cascade.
  See \cite{Strassler:2005qs} for further discussion on the
weak coupling version of the cascading theory.
   We have bifundamental fields $A_i$ and $B_a^\dagger $ and their complex conjugates, which
   are anti-bifundamentals.
 The classical baryonic branch has two regions one with $B_a=0$ and one with $A_i=0$.
 Here we concentrate on the first and set $B_a=0$ and
  $ A_i = C \,  \Phi_i \otimes {\bf 1}_{M }  $ where $C$ is an arbitrary complex constant and
  $\Phi_i$ are the following two  $k\times (k+1) $ matrices
\bea \label{matrices}
\Phi_1 =
\left(
\begin{array}{cccccc}
\sqrt{k} & 0 & 0 & \cdot & 0 & 0 \\
0 & \sqrt{k-1} & 0  & \cdot & 0 & 0\\
0 & 0 & \sqrt{k-2} &  \cdot & 0 & 0\\
\cdot & \cdot & \cdot & \cdot & \cdot\\
0 & 0 & 0 & \cdot & 1 & 0
\end{array}
\right)
~~~~\Phi_2 =
\left(
\begin{array}{cccccc}
0 &  1 & 0 &  \cdot & 0 & 0 \\
 0 & 0 & \sqrt{2} &  \cdot & 0 & 0\\
0 & 0 & 0 & \sqrt{3} &  \cdot & 0 \\
\cdot & \cdot & \cdot & \cdot & \cdot\\
0 & 0 & 0 & 0 & \cdot & \sqrt{k}
\end{array}
\right)
\eea
We can view this as a solution in the $SU(k) \times SU(k+1)$ quiver theory and we recover the solution of
the $SU(kM) \times SU((k+1)M)$ theory by multiplying each entry by the $M\times M$ identity matrix, ${\bf 1}_{M}$.
Thus we see that we can set $M=1$ for the time being and we will restore the $M$ dependence at the end.

Setting $B_a=0$ the   D-term equations of the theory are the following
\bea
A_1 A_1^\dagger + A_2 A_2^\dagger  & = &   (k+1) |C|^2 \mathbf{1}_k ~,\nn\\
 A_1^\dagger A_1 + A_2^\dagger A_2   & = & k |C|^2  \mathbf{1}_{k+1}~.
 \label{dterms}
\eea
The constant $C$
 is a modulus of the solution since
the $D$ term constraints set to zero only  the $SU(N)$ part of \nref{dterms}.
 The fact that the constants in the
two lines of \nref{dterms} are related   follows from taking the trace on both sides.
 We have defined $C$ in such a way that the expectation value of the scalar operator ${\cal U}$,
 the ${\cal N}=1$ partner of the baryonic current, is\footnote{ Our normalization of
  ${\cal U}$ differs from the one in \cite{DKS} by a factor $M$. In our normalization the baryon
  operator, ${\cal A}$ has charge one under the baryonic current.}
 \be \label{udefi}
  { \cal U } = { 1 \over M k (k+1) } \mathrm{Tr}[ A_i^\dagger A_i - B_i^\dagger  B_i ]  \sim   |C|^2~.
\ee
If we {\it were }  to gauge the baryon current,
then  $C$ would be the FI term we would need to have the above  VEVs.
 However, we are {\it not} gauging the baryon number current. The factors of $k$ were introduced in  \nref{udefi} because we want
 to normalize the baryon number current so that the bifundamental $A_i $ has charge ${1 \over M k(k+1) } $ after
$k$ steps in the cascade \cite{Aharony:1995qs}.
More explicily, the baryon operators have the schematic form ${ \cal A} \sim ( A_i)^{ k(k+1) M } $, with the precise index
contractions given in
 \cite{Aharony:2000pp}. We want these to have   baryon charge one.
  However, since we do not know the K\"ahler potential, we do not know
if \nref{udefi} will remain as the proper expression for the operator as we make the coupling stronger. In general
${\cal U}$ is defined as the partner of the baryon current.
For the moment we will just do
the computation in a weakly coupled theory for fixed $k$.

We now define the following $k\times k$ matrices
\bea
  L_1  &=&  { 1 \over 2} (\Phi_1 \Phi_2^\dagger + \Phi_2 \Phi_1^\dagger ) \nn\\
 L_2  &=&  {i \over 2} (\Phi_1 \Phi_2^\dagger - \Phi_2 \Phi_1^\dagger ) \label{sutwoleft}
 \\
  L_3 & =&   { 1 \over 2} (\Phi_1 \Phi_1^\dagger - \Phi_2 \Phi_2^\dagger) \nn
  \\
  & & ~~{\rm and} ~~~  \Phi_1 \Phi_1^\dagger + \Phi_2 \Phi_2^\dagger = (k+1) {\bf 1}_k  \nn
\eea
We find that the hermitian matrices $L_i$ obey the $SU(2)$ commutations relations.  In addition
 we find that the Casimir operator is
\bea
L_1 L_1^\dagger  + L_2 L_2^\dagger  + L_3 L_3^\dagger  = \frac{1}{4}(k^2 -1) \mathbf{1}_{k}
\eea
Thus we have the spin $j = { k-1 \over 2}$, or $k$ dimensional, irreducible
representation of $SU(2)$.
We can do the same for the matrices multiplied in the other order. We define
\bea
 R_1  &=&  { 1 \over 2} ( \Phi_1^\dagger \Phi_2 +  \Phi_2^\dagger \Phi_1 )\nn\\
  R_2  &=&  { i \over 2} ( \Phi_2^\dagger \Phi_1  - \Phi_1^\dagger \Phi_2 )  \label{sutworight} \\
  R_3  &=&  { 1 \over 2} ( \Phi_1^\dagger \Phi_1 -  \Phi_2^\dagger \Phi_2 ) \nn  \\
  & & ~~~ {\rm and } ~~~
   \Phi_1^\dagger \Phi_1 +  \Phi_2^\dagger \Phi_2  = k {\bf 1 }_{k+1}\quad \nn
   \eea
 In this case the Casimir is
\bea
R_1 R_1^\dagger + R_2 R_2^\dagger + R_3 R_3^\dagger = \frac{1}{4}k (k+2) \mathbf{1}_{k+1}
\eea
thus it is a spin $j = {k\over 2 } $, or $(k+1)$-dimensional,  irreducible representation of $SU(2)$.

The commutation relations of these matrices with the $\Phi_i$ show that the $\Phi_i$ transform in the
fundamental representation of the sum of the two $SU(2)$ groups.
This is important for arguing that the $SU(2)$ global symmetry that acts on the $i$ index of $A_i$ is unbroken,
once we combine it with appropriate gauge transformations. The appropriate gauge transformations
are generated by $SU(2)$ matrices living in each of the gauge groups. These matrices are the matrices
$L_i$ and $R_i$ discussed above. This unbroken $SU(2)$ symmetry is important for classifying fluctuations around the background.

In fact, the matrices \nref{matrices} really define a fuzzy supersphere,  \cite{Pais:1975hu,Grosse:1995pr} or
\cite{Balachandran:2005ew,Ydri:2007tz} for reviews. In addition to the ``even'' generators
$L_i$ and $R_i$ we also have ``odd'' generators  given by the $\Phi_i$.
 We did not find the supersphere perspective useful for what we do here, but it is a curious fact.

\subsection{The spectrum of quadratic fluctuations}

In order to demonstrate the emergence of a two-sphere we want to show that the quadratic fluctuations
around this vacuum behave as Kaluza-Klein modes on a two-sphere.

We start considering the four dimensional gauge fields. Most of them are higgsed along the baryonic branch.
We want to show that the spectrum of massive gauge fields agrees with what one expects from a Kaluza-Klein
compactification of a six dimensional gauge theory on a two-sphere.
In other words, when we set a non-zero VEV for the fields $A_i$ we are higgsing the
$SU(k) \times SU(k+1)$ gauge fields.  We denote the four
dimensional gauge fields as $a^L_\mu$ and $a^R_\mu$. They are in the adjoint of $SU(k)$ and $SU(k+1)$ respectively.
The Higgs fields   mix $a^L$ and $a^R$. In order to avoid unnecessary notational clutter, we sometimes drop
the four dimensional index $\mu$. Alternatively we concentrate just on one particular component of the four dimensional gauge
field.
The masses for the gauge bosons come from expanding the kinetic term for the Higgs fields $A_i$
\be
\sum_i \mathrm{Tr}[  D_\mu A_i^\dagger D^\mu A_i ]  ~,~~~~~~~~~~D_\mu A_i = \partial_\mu A_i + i ( a^L_\mu A_i - A_i a^R_\mu )~.
\ee
We get the structure $\mathrm{Tr} [ ( a^L \Phi_i - \Phi_i a^R )^\dagger   ( a^L \Phi_i - \Phi_i a^R )] $.
Expanding this out we get
\bea
 \mathrm{Tr} \left[ (k+1)(a^L)^2  + k (a^R)^2   -  2 \Phi_1^\dagger a^L \Phi_1 a^R - 2  \Phi_2^\dagger a^L
\Phi_2 a^R \right] ~.
\label{relevant}
\eea

We will now expand $a^L$ and $a^R$ in fuzzy spherical harmonics,
\be
 a^L \sim \sum_{l=0}^{k-1} c_l (L)^l  ~~~~~~~~~~a^R \sim \sum_{l=0}^k  c_l (R)^l
 \ee
 where the $(L)^l = L_{(i_1} \cdots L_{i_l)}   - $traces. These are simply products of
 the matrices introduced in \nref{sutwoleft}. They transform in the spin $l$ representation of
 the unbroken $SU(2)$. The coefficients $c_l$ have the
 corresponding indices, which are the same as the indices of ordinary spherical harmonics.
 $SU(2)$ symmetry allows us to decouple different values of $l$, but since we have
 $a^L$ and $a^R$ we end up with a two by two matrix.
 In order to compute this  matrix, we need to define an operator $S$ via
 \be
S ( M) \, =\,  \Phi_1 M \Phi_1^\dagger +  \Phi_2 M  \Phi_2^\dagger~.
\label{spinchange}
\ee
We think of $S$ as an operator which sends $(k+1) \times (k+1)$ matrices into $k\times k$ matrices and
it commutes with the unbroken $SU(2)$. It is easy to check that
\bea
S ( R_+^l ) &  = & (k+ 1 + l ) L_+^l
\label{changespin}
\eea
where $R_+= \Phi_1^\dagger \Phi_2$ and $L_+ = \Phi_2 \Phi_1^\dagger$ are the raising generators
\nref{sutwoleft}, \nref{sutworight}.
Since the action of $S$ respects $SU(2)$, it means that it acts in
this way on any of the elements of $R^l$ transforming according to the $l$th spherical harmonic.
We can similarly define an operator $\tilde S$ as
\bea
\tilde S  ( M) \, &= &\,  \Phi_1^\dagger M \Phi_1 +  \Phi_2^\dagger M \Phi_2~.
\label{spinback}
\eea
For this we have
\bea
\tilde S (L_+^l) &  = & (k - l ) R_+^l ~.
\label{backspin}
\eea
Using \nref{backspin},  \nref{changespin} and  \nref{relevant} we get the
following two by two matrix for each value of $l$
\bea
\lambda \left( \begin{array}{c} a^L_l \\ a^R_l \end{array} \right) = \left(
\begin{array}{cc}
k+1  & - (k+1+l)\\
-  (k-l)  &  k
\end{array}
\right)  \left( \begin{array}{c} a^L_l \\ a^R_l \end{array} \right)
\label{massmatrix}
\eea
with eigenvalues
\bea
\lambda_{l, \pm} & = & k + \frac{1}{2}  \pm \sqrt{\left(k+\frac{1}{2}\right)^2 -  l (l+1)}~.
\label{gaugeigen}
\eea
This formula is valid for $l=0,1,\cdots k-1$. For $l=k$ we have that $a^L=0$ and the eigenvalue is simply $k$, which
is the same as $\lambda_-$.
 \nref{gaugeigen}  gives   a zero eigenvalue for   $l=0$. This gives the unbroken gauge symmetry.
In the case that we have an $SU(k) \times SU(k+1)$ group there is no unbroken gauge symmetry and this mode is
not present. On the other hand, if we have $SU(k M) \times SU((k+1)M)$ then this mode gives the unbroken $SU(M)$ gauge
symmetry. This corresponds to the $SU(M)$ gauge symmetry on $M$ fivebranes.
Let us expand the eigenvalues for large $k$ and fixed $l$. We get
\bea
\lambda_{l,-} & = & \frac{l(l+1)}{2k +1} +\dots   ~,~~~~~~~~~~~~~~l\ll k~, \nn\\
\lambda_{l,+} & = & 2k+1 - \frac{l(l+1)}{2k +1}+ \dots  ~,~~~~~~~~~l\ll k~. \label{eigenv}
\eea
The lower values, $\lambda_{l,-}$,
 agree with our interpretation in terms of Kaluza-Klein modes on the two-sphere.
The other ones, $\lambda_{l,+}$ are large, as long as $k$ is large, and we will not offer any interpretation for them.
We view them as a UV effect associated to the precise fashion in which this fuzzy sphere is approximating the ordinary sphere.
It would be nice to see if they have a simple physical interpretation.
\begin{figure}[ht!]
\epsfxsize = 8cm
\centerline{\epsfbox{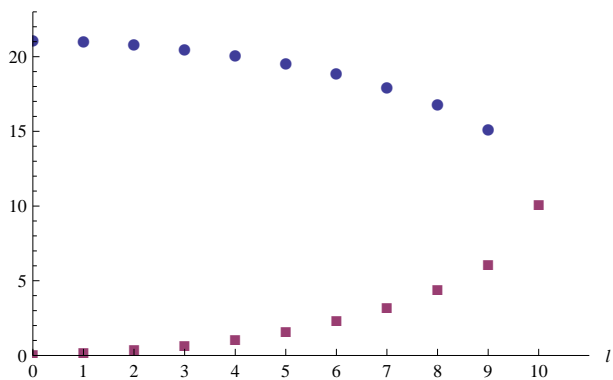}}
\caption{The eigenvalues $\lambda_{+}$ (upper) and $\lambda_{-}$ (lower) as functions of $l$, for $k=10$.}
\label{eigenfig}
\end{figure}

So far we have shown that the four dimensional components of the gauge field correctly reproduce
what we would expect from Kaluza-Klein reducing a six dimensional theory on a two-sphere.
One could do a similar analysis expanding the fields $A_i$ around their vacuum expectation values in
terms of the fuzzy sphere spherical harmonics. These get a mass from the D-terms potential. However, it
is not necessary to do this explicitly,
 since  ${\cal N}=1$ supersymmetry implies that there should be a scalar partner with the same masses
as the ones from \nref{eigenv}. In fact, this is enough to account for all the modes that come from $A_i$.
The counting is the following. We have $ 2 \times 2 \times k (k+1)$ real components for the two complex
fields $A_i$. On the other hand, the number of massive vector fields is $k^2 -1  + (k+1)^2 -1 = 2 k (k+1) -1$.
Thus, we see that all components of $A_i$ are involved in the ${ \cal N} =1$
Higgsing of the gauge bosons except for a single
complex field, which is simply the field $C$.

In addition we can consider the Kaluza-Klein modes of the $B_a$ fields.
Here we can compute their masses directly from the superpotential
\be
W = h \mathrm{Tr} [ \epsilon^{ab} \epsilon^{ij} A_i B_a A_{j} B_{b} ]~.
\ee
Again we now expand $B_a$ into spinor spherical harmonics. The reason we get spinors is that we get
an odd number of fields $\Phi_i$ or $\Phi^\dagger_i$. So we have $B_{a} = \sum_{l=0}^{k-1}  b_a^l X^l$
were $X^l \sim  \Phi^i , ~\Phi^{(i_1\dagger} \Phi^{i_2} \Phi^{i_3)\dagger} , \cdots $, for $l=0,1,\cdots $.

The eigenvalues of the superpotential are then $\pm (l+1)$ and the spin under the unbroken $SU(2)$
 is $j = l + {1 \over 2}$.
This agrees with the spectrum of the Dirac operator on the fuzzy sphere \cite{Grosse:1995pr}.
The structure of the superpotential is then
\be
W = h \epsilon_{ab} \sum_{l=0}^{k-1}  \langle b^l_{a} b^{l}_b \rangle~,
\ee
where the angle brackets denote the antisymmetric $SU(2)$ invariant inner product of two
SU(2) representations with half integer spin. The mass eigenvalues come from the F-term potential
$\sum_{a=1}^2|\de W /\ \de B_a|^2$, thus in the end we have that $\lambda_{l,B} \,= \,|C|^4 h^2 (l+1)^2$.

\subsection{Fuzzy sphere parameters}

We found three types of fields, each organized into $SU(2)$ representations.
Two towers of vectors from the gauge fields
$a^L$, $a^R$, this tower also contains the ${\cal N}=1$ scalar partners of the massive gauge fields coming
from part of the fluctuations of $A_i$.
Finally we have one tower    from the fields $B_a$. The spectrum is  summarized in the following table
\begin{table}[ht!]
\centerline{
\begin{tabular}{|c|c|c|c|}
\hline
fields &  $SU(2)$ spin  & \# superfields   & eigenvalues \\
\hline
$ a^L_\mu ~, a^R_\mu $  + scalar($\delta A_i$) & $ j = l$  &
$ 1$  &  $ \lambda_{l,-}, \lambda_{l,+} $ \\
\hline
$B_a$  &   $j = l + { 1 \over 2} $ & $ 2 $ & $ \lambda_{l,B} $ \\
\hline
\end{tabular} }
\label{spectrumta}
\end{table}

For $l \ll k$ the eigenvalues are
\bea
\lambda_{l,-} \,\sim  \,\frac{g^2|C|^2}{2k+1} l(l+1)~, \quad
\lambda_{l,+}  \,\sim  \, g^2|C|^2 (2k+1)~, \quad
\lambda_{l,B} \,= \,|C|^4 h^2 (l+1)^2~.
\label{sumeigen}
\eea
The dependence on the gauge coupling
 comes from the fact that we normalize the
YM term without a coupling constant by rescaling the gauge fields by $g$. We have set the two $g$'s equal when
we derived the spectrum.  Of course, the couplings would run in opposite directions, and we would
get a slightly more complicated expression. The correct formula for $\lambda_{l,-}$, for
small values of $l$,  is as in \nref{sumeigen}
but with ${ 1 \over g^{2}}  \to { 1 \over g_+^2} \equiv { 1 \over g_L^2 } + { 1 \over g_R^2 }$.

Neglecting the highly massive states with $\lambda_{l,+} \sim k$, we see that the
spectrum is very simlar to  the spectrum of the
fivebrane theory compactified on an $S^2$, as computed in \cite{Andrews:2006aw} (see pages 21-22 in
 \cite{Andrews:2006aw}).
 The only difference is that in \cite{Andrews:2006aw}
 the modes with half integer spin had masses which were set by the same
 overall scale as the modes with integer spin. Here the ratio of their masses
  involves $h^2 |C|^2$ which is an arbitrary
 parameter.
 {}From the gravity  dual that  we discussed in the previous sections we would have expected these
 modes to have the same mass, as in \cite{Andrews:2006aw}.
 We should not be surprised by this mismatch, the field theory computation we did here was  for a weakly
coupled theory. At strong coupling we expect that the coefficient of the superpotential should be
determined by the other parameters. Further discussion can be found in \cite{Strassler:2005qs}.

{}From the expression of the masses of the four dimensional gauge fields we can read off the radius
of the fuzzy sphere as well as the non-commutativity parameter
\bea
{ 1 \over |C|^2 R^2_{\rm Fuzzy} } = { 1 \over \langle {\cal  U} \rangle  R^2_{\rm Fuzzy} } \,
 \propto \,  {g^2_+ \over k  } ~,~~~~~~~~~~~~~~\theta_{\rm Fuzzy} \propto  { 1\over k }
\label{defradius}
\eea
were $|C|^2 R^2_{\rm Fuzzy}$ is the radius of the sphere in units of $|C|^{-2}$, or the VEV of the operator
${\cal U}$. This is setting the scale of the overall mass of the gauge bosons and it is the natural scale to use.

So far, we are not finding any relation between $|C|$ and $k$. And indeed there is no  relation in
the weakly coupled field theory. However, once we include the effects of the cascade we expect $k$ and
$|C| $ to be related. For a given $|C|$, or a given VEV of the gauge invariant operator ${\cal U}$ \nref{udefi},
we should find the value of $k$ corresponding to the appropriate region of the cascade. As we increase $|C|$ we see that
$k$ should increase. We know that the running of the difference between the couplings goes like
$ 8 \pi^2 ( { 1 \over g_L^2} - { 1 \over g_R^2 } ) \sim 3 M \ell  + {\rm constant} $,
 where $\ell = \log ({\rm scale})$ is the RG time \cite{KS}.
 The amount of ``time'' or $\Delta \ell$ for each step in the cascade can be calculated by setting $g_L =\infty$ and then
 see how much we should run until $g_R \to \infty$. This gives $\Delta \ell_{\rm 1-step} = 8 \pi^2/( 3 M g_+^2 )$.
 The natural scale is here set by the value of the VEVs which is in turn given by $|C|$. Thus the amount of RG time from the
 IR scale $\Lambda$ to the scale $|C|$ is given by $\Delta \ell \sim \log { |C| \over \Lambda } $.
Then the number of steps in the cascade goes as
\be
 k \sim { \Delta \ell \over \Delta \ell_{\rm 1-step} } \sim
 { 3 M g_+^2 \over 8 \pi^2  } \log { |C| \over \Lambda } = { 3 M g_+^2 \over 16 \pi^2  } \log { \langle {\cal U} \rangle \over \Lambda^2 }  ~,~~~~~~~
 { 1 \over g_+^2} = { 1 \over g_L^2} + { 1 \over g_R^2}
 \ee
 where $\Lambda$ is the scale of the last step of the cascade in the IR. $k$ is telling us how many steps away we are from the
 last step of the cascade. Of course, $k$ is an integer while the right hand side is a continuous variable.
 Here we are considering the large $k$ limit where the distinction is not important.

 For large values of $|C|$,  it is natural to measure the size of the $S^2$ in units of the
 VEV of the  operator ${\cal U}$ which has dimension two. This simply gives from \nref{defradius}
 \be \label{finans}
\langle {\cal U } \rangle   R^2_{\rm Fuzzy}
 \propto { k  \over g_+^2 } \sim { 3 M \over   16 \pi^2 }  \log { \langle { \cal U}\rangle \over  \Lambda_0^2 } + \cdots
  \ee
where $\Lambda_0$ is the scale at the last step of the cascade, normalized with the factor of $M$ natural
from the 't Hooft counting point of view\footnote{In this normalization the gaugino bilinear expectation
value is $\langle \mathrm{Tr} [ \psi^2] \rangle \propto  M (g^2_+ M) \Lambda_0^3 $, see \nref{tensionKS}. We have ignored a factor of $M^2 g_+^{2/3}$ inside the $\log $ in \nref{finans}.}.

\subsection{Comparison with the gravity picture}

We can now compare to the quantities that we had in the gravity analysis.
First we recall that the VEV of the field ${\cal U}$ is proportional to \cite{DKS}
\be \label{expectugra}
\langle {\cal U} \rangle   \propto M  U \Lambda_0^2 \propto M  e^{ 2 t_{\infty} \over 3 } \Lambda_0^2~.
\ee
We have seen that the metric in this solution is basically the resolved
 conifold with $M$ fivebranes
 wrapping it, plus a large amount of D3 brane flux.
 The amount of D3 brane flux that we have on the fivebranes can be
 determined by computing the value of the $B^{NS}$ field on the two-cycle near the tip of the resolved
 conifold. The value of the $B^{NS}$ field only varies logarithmically, so it does not matter precisely where
 we evaluate it, as long as it is around $t \sim t_\infty$, which is the region where the metric looks
 like that of the resolved conifold. In fact we have
\be \label{bfi}
\left. { 1 \over (2 \pi)^2 \alpha'} \int_{S^2}  B^{NS} \right|_{t = t_\infty}  \propto  {   g_s \over 2 \pi  }   M t_{\infty} = k~.
\ee
We have identified this with the number of steps from the bottom
of the cascade in the gravity approximation, since
it gives us how many D3 branes we have dissolved on the D5 branes:  $N_3 = k M$.

We would also like to have some way of estimating the size of the $S^2$ on which we put the fivebranes.
We see that the radius of the $S^2$ of the conifold is proportional to  $ t_\infty$
before we do the boosting procedure \nref{diff}.
The boosting procedure introduces the warp factor $ \hat h $
which multiplies the spacetime direction and a similar
factor multiplying the spatial directions \nref{KSmetricfi}.
The radius of the $S^2$ is then given by
\be \label{rclosed}
 r^2_{S^2} \propto  \hat h^{1/2} g_s M t_\infty~.
 \ee
 Note that in this region the dilaton is constant and $e^{ 2 (\phi -  \phi_\infty)} \sim 1 $.
 We find that the fivebrane has a large amount of $B$ field and we
 should use the appropriate expression for the open string metric on the fivebranes.
We use
 the formulas for the open string metric on branes when we have
 a large $B$ field in eqn (2.5) in  \cite{SWnoncomm}
\be \label{swformula}
G^{ij}_{\rm open} \sim { r^2_{\rm closed}  \over  B^2 } ~,~~~~~~\theta^{ij} \sim { 1 \over B} ~,~~~~~{\rm for}
~~~~~r^2_{\rm closed} \ll B
\ee
where $r$ is the radius of the closed string metric, which appears in \nref{rclosed}.
We face the problem that $\hat  h$ diverges where the branes are sitting, but of course, this is already taking
 into account the backreaction of the branes. We should really evaluate $\hat h$ at some distance from the
 point where the fivebranes are sitting. It turns out that the final answer \nref{finans} does
  not depend on $ \hat h$. However,
 we see that as we are in the region that the metric is accurately given by the resolved conifold, but away
 from the origin of the resolved conifold   (say at $\rho/\alpha \gg 1/t_\infty$ but $\rho \sim \alpha$  in \nref{RCmetric}),  we get that $\hat h $ is becoming very
 small as $ U \to \infty$.   In fact, we have
that $r_{\rm closed}^2 \propto \sqrt{ t_{\infty} }$ vs. $B \propto t_\infty$. See around eq. \nref{ptwarp}.
For large $t_\infty$,  this justifies the use of \nref{swformula}.

 {}From \nref{bfi} and \nref{swformula} we see that
 the non-commutativity parameter is indeed as in the fuzzy sphere
 construction \nref{defradius}.
 Similarly, we can compare the masses of the Kaluza-Klein modes on the fivebrane.
 These masses are proportional to
 \be
 m_{KK}^2  = [ \hat h ^{-1/2} U M g_s \Lambda_0^2 ] { l (l+1) \over r^2_{\rm open} } =
  l ( l+1)  \hat h ^{-1/2}  U M \Lambda_0^2 { r^2_{\rm closed} \over B^2 } \sim l (l+1) { U \over t_\infty} \Lambda_0^2
 \ee
 where the factor of $[\hat h^{-1/2} U M g_s \Lambda_0^2 ]$ comes from the warping of the four dimensional space
 in \nref{KSmetricfi} and
 we have used that $r^2_{\rm closed}$ is $r^2_{S^2}$ computed
 in \nref{rclosed}.   $l$ labels the angular momentum on $S^2$.
 We see that once we express these modes in units of the expectation value of ${\cal U}$ from
 \nref{expectugra}  we get
\be \label{finansfi}
{ m_{KK}^2 \over \langle {\cal U } \rangle } \propto  { m_{KK}^2 \over M U \Lambda_0^2 }
 \propto   l (l+1)     { 1  \over M  t_\infty }     \propto { l (l+1) } { g_s  \over k }
  ~,~~~~~~g_s \propto g_+^2
\ee
Thus we see that we get agreement also for the $k$ dependence of the
 radius of the fuzzy sphere in \nref{defradius}.

Notice that the dilaton $\hat \phi$ is very close to a constant up to the region that is
very close to the branes. Once we analyze in the near brane region we see that the dilaton
$\hat \phi = - \phi$ starts decreasing rapidly, see \nref{muval} in appendix A. This implies,
due to \nref{f3rrsec}, that $H_{NS} \to 0$   (for large $t_\infty$)
rapidly and the solution becomes very similar to the
straight S-dual of \nref{ansatz}. This is related to the fact that the effects of non-commutativity
on the fivebrane become less important as we go to the IR.

Notice that the emergence of the fuzzy sphere relied on the VEVs  for $A_1$ and $A_2$ given by \nref{matrices}, which
are a solution of the D-term equations. On the other hand, we did not rely on the details of the superpotential. By setting
$B_a =0$ we ensured that $ \partial W =0$. In particular, we know that the Klebanov-Strassler theory could arise from
an ${\cal N}=2$  $\Z_2$ orbifold of ${\cal N }=4$ plus a mass deformation \cite{KW}.
Since the mass deformation only enters in the
superpotential, we can easily check that the VEVs for $A_i$ in \nref{matrices} continue to be good vacua. Thus, when
these VEVs are much larger than the mass we expect that the configuration should have a description in terms of
a $SU( M (k+1) ) \times SU( M k)$  $ {\cal N}=2$ quiver theory, see \cite{Aharony:2000pp} for further discussion on
this theory. In this case,  we expect that the proper description
of the vacuum should be in terms of  fivebranes that are wrapping the $S^2$ of an Eguchi-Hanson space.
Solutions corresponding to such configurations were presented in \cite{ehsolutions}.
However, we did  not check the details.

\section{Discussion}

In this paper we have analyzed various solutions describing closely related
configurations. The solutions are not new, and they are contained in
\cite{Butti}. Nevertheless, we think that the points we have made are not
generally appreciated.

First, we have discussed the most basic solution from which all others follows.
This is the solution for a number of fivebranes wrapping the $S^2$ of the resolved conifold.
Alternatively we can just as well say that it is the solution describing a deformed
conifold with flux. In both cases, the geometry is not that of the resolved or deformed conifold.
The solution interpolates between the deformed conifold with flux and the resolved conifold with branes and
it is a simple realization of the geometric transition described in \cite{vafageometric}.
With NS three form field strength the four dimensional string metric is unwarped, which justifies the
first word in our title.

 The solution we discussed is also one of the few explicit examples of torsional geometries, in the sense
of \cite{Strominger:1986uh,Hull:1986iu}. In particular, the geometry is complex,
 but not K\"ahler. Thus, this geometry can be viewed as a non-K\"ahler version of the conifold,
 where the metric is not Ricci-flat\footnote{The CV-MN solution is  also a
  non-compact,  non-K\"ahler geometry. But it asymptotes to a linear dilaton background.}.
 The solution discussed here may be  describing
a  region of a bigger compact  manifold.
One difference with the conifold is that no cycle goes to zero size, and the geometry is always smooth.
 It  is  a natural arena for studying aspects of non-K\"ahler geometry.

Starting from this solution one can add D3 brane charge by a certain U-duality
transformation.  This gives a useful perspective on  the  solution
of \cite{Butti}, representing the gravity dual of the baryonic branch of the Klebanov-Strassler theory.
 In fact, this could have been another avenue for deriving that solution.
 The BPS equations become  simpler  with only non-trivial $H_3$ and dilaton.
  It would be nice to see if other explicit interesting
solutions can be constructed in this way, starting from solutions of Type I supergravity.

We have also discussed the interpretation of the U-duality transformation in a T-dual brane picture. In this context
the duality corresponds to a simple \emph{rotation} of the NS branes. Various features of the supergravity
solutions may be then understood heuristically  from this picture.

One basic lesson of this analysis is that going far along a baryonic branch in confining theories with fractional
branes is related to resolving the singularity and wrapping some branes on the resulting two-cycles. This picture
could be particularly useful for cases where one cannot find the explicit solutions. One interesting case would be
the theory studied in \cite{susybog} which is supposed to display a runaway behavior \cite{Intriligator:2005aw}
pushing it far along the baryonic
branch. Thus, it might be possible to find the gravity picture of the runaway
behavior. One would   start  with a suitably resolved
space\footnote{Several explicit Ricci-flat K\"ahler metrics on  (partially) resolved Calabi-Yau singularities were presented in
\cite{Martelli:2007pv}.},  add  branes, and presumably find that   there is residual force pushing
the branes away,  as opposed to the case in
this paper where we have an exact modulus.

The emergence of a two-sphere when we go along the baryonic branch is another observation that
we have made. We have seen that the scaling of the size of the sphere matches quite well between the field theory
and the gravity description. The fact that the fuzzy sphere arises does not depend too much on the details of the theory.
It only relied on the existence of a quiver description with two different ranks. It would be nice to explore this
phenomenon in more generality by considering a general class of theories. One closely related example is the
picture for vacua of the massive ABJM theory discussed in \cite{Nastase:2009ny}.

The analysis in this paper is probably also useful for studying in more detail the inflationary model
proposed in \cite{McAllister:2008hb} which involves wrapping fivebranes on the $S^2$ of the conifold.

\subsection*{Acknowledgments}

We are very grateful to I. Klebanov, C. Nu\~nez, N. Seiberg and Y. Tachikawa
for discussions. The results of this paper were obtained whilst D.M. was Member of
the Institute for Advanced Study, and was supported by NSF grant PHY-0503584.
This work was supported in part by U.S.~Department of Energy grant \#DE-FG02-90ER40542.

\appendix

\section{More details  on the solution}
\label{remarks}

In this a appendix we consider the equations \nref{equationf}\nref{equationc}, which we reproduce here\footnote{The function $c$ used here is
  related to $a$  in \cite{Butti} by
$ c = - { a (\sinh t - t \cosh t ) \over ( 1+ a \cosh t ) } $. }
 \bea
 f ' &=& 4 \sinh^2 t \, c\label{equationfa}
 \\
 c' &=& { 1 \over f } [ c^2 \sinh^2 t - (t \cosh t -  \sinh t )^2 ]~.
 \label{equationca}
 \eea
We will collect a few facts about these equations.
We can define a new variable $\tau$ via $d\tau = \sinh^2 t dt$. Then the equations become
\bea
 \partial_\tau f   &=& 4  c  \label{equationftau}
 \\
 \partial_\tau  c  &=& { 1 \over f } [ c^2  -k(\tau) ]    ~,~~~~~~~k(\tau) = \left( { t \over \tanh t} -1\right)^2 \label{equationctau}
 \\
 \tau &=&   { 1 \over 2 } ( \cosh t \sinh t - t) \label{tauexp}
 \eea

The equations \nref{equationftau}, \nref{equationctau}
 can be written as a second order equation for $f$,
\be
 f \partial_\tau^2 f =  { ( \partial_\tau f )^2 \over 4} - 4 k(\tau)
\label{tauode}
\ee
which is  the equation of motion  for the action
\be \label{lagra}
S  =  \int d \tau \left[  { 1 \over 16 } {  ( \partial_\tau f )^2 \over \sqrt{f} } +   { k(\tau) \over \sqrt{f} } \right]~.
\ee
We could also introduce a new variable $x = f^{3/4}$. Then the lagrangian has the form $\dot x^2 + k(\tau) x^{ -2/3}$.
This is a negative potential. The particle starts at $x=0$ at $\tau =0$ and the rolls off down the potential
as $\tau \to \infty$.

The Hamiltonian associated to the lagrangian in \nref{lagra} is not conserved, and it is given by
\be \label{hamiltonian}
H ={ 1 \over 16 }  { (\partial_\tau f )^2 \over \sqrt{f} } - { k(\tau) \over \sqrt{f} }
=  e^{ 2 (\phi - \phi_0) }
\ee
where we noted that the  Hamiltonian is equal to  the dilaton in \nref{dilatonvalue}.
Using   \nref{smallt}, \nref{larget} we see that  this has the following values at $t=0$ and $t=\infty$
\be
H(0) = \gamma^3 ~,~~~~~~~~~~~H(\infty) = { 1 \over 9 } e^{ - t_\infty} = { 8 \over \sqrt{3} } U^{-3/2}~.
\ee
The derivative of the Hamiltonian on a solution  is given by the explicit time dependence of the Lagrangian
\be \label{ennoncon}
\partial_\tau H = - \partial_\tau L =   -
{(\partial_\tau k)\over \sqrt{f}}~.
\ee
This  is negative  since $k(\tau)$ is an increasing function of $\tau$ \nref{equationctau}.
 Thus the dilaton is a maximum at $t=0$ and
it then decreases monotonically as $\tau \to \infty$.
In fact, for large times the change in the energy goes to zero due to \nref{larget}. In fact, we can
compute the first subleading term for large $t$ which has the form
\be \label{dilsub}
\hat h(t) = { H(t) \over H(\infty) } -1 =
2 \times  3^3 \, t \, e^{ - { 4 \over 3 } t }\,  e^{ { 4 \over 3 }  t_\infty} = U^2 {3  t \over 8 } e^{ - 4/3 t}~.
\ee
This function appears in  the expression of the boosted solution and also in the solution with
Klebanov-Strassler asymptotics \nref{KSasymp}. This overall factor of $U^2$ cancels out in \nref{KSasymp}.

\subsection{The solution for small $U$, or $t_\infty \ll 0$}

When $U$ is small and $t_\infty$ is very negative, then we have that $\gamma \gg 1$. In this case
the particle described by \nref{lagra} moves very quickly to large values of $f$ where the derivative of
the energy becomes very small. Thus, in this regime the energy is conserved to first approximation and
the dilaton is constant.
In the limit that $\gamma$ is very large we can find an approximate solution to these equations
by neglecting   $k(\tau)$ in \nref{equationctau}. This approximate solution has the form
\be \label{defcon}
c^3 = \gamma^6 \, 3 \, \tau  ~,~~~~~~~~ f = \gamma^{-6} c^4 ~,~~~~~~~~~~~~~ \gamma \gg 1
\ee
with $\tau $ in \nref{tauexp}.
This solution, inserted in the ansatz, gives the deformed conifold metric. More precisely, in a scaling
limit where $\gamma^2 \to \infty$, and up to an overall scale $\gamma^2$ in the metric,
 we get precisely
the deformed conifold, \nref{DCmetric}. For large and finite $\gamma$, we get a metric which is very close to the deformed
conifold, but in addition we have a non-vanishing three form NS flux on the $S^3$ of the deformed conifold.
The relation between $\gamma$ and $U$ in this regime is
\be \label{gamla}
{ U \over 12 } = e^{ 2 t_\infty \over 3 } \sim { 1 \over 3^{4/3} \gamma^2 }
\ee
which can be obtained comparing the large $t$ behavior of \nref{defcon} and \nref{larget}.

We can also  find the subleading correction to the dilaton
 by
using the energy non-conservation equation \nref{ennoncon}
\be \label{subleadingdil}
 \hat h = e^{ 2( \phi - \phi_\infty ) } -1 =   { 1 \over \gamma^3 } \int^\infty_{t} { \partial_t k \over \sqrt{f} } =
  { 1 \over \gamma^4} 2^{4/3} 3^{-2/3} \int^\infty_t  dt' { ( t' \coth t' -1) \over \sinh^2 t'}
( \sinh 2 t' -2 t')^{1/3}
\ee
This expression is necessary to recover  the Klebanov-Strassler  limit of the solution \nref{KSasymp}.
Of course, the large $t$ limit of \nref{subleadingdil} agrees with the general expression \nref{dilsub}, after using \nref{gamla}.

\subsection{The solution for large $U$, or $t_\infty\gg 0$}

 We now want to study the solution in the regime $\gamma \sim 1$.
 For $\gamma =1$ we have the
  CV-MN \cite{volkovone,volkovtwo,MN} solution
\be
\label{cvmn} c = t ~, ~~~~~~ f   =  t^2 \sinh^2 t  - ( t \cosh t -
\sinh t )^2 =  t ( \sinh( 2t) - t) - \sinh^2 t~. \ee
This solution
does not go over the conifold at infinity. However, as soon as
  $1 < \gamma $, the asymptotic form of the solution at
large $t$ changes and it becomes that of the  conifold.
The solution stays very close to \nref{cvmn} up to a large value of $t$ and then it
starts deviating from it.
The large $t$ form of \nref{cvmn} is
\be\label{cvmnlarge}
 c = t ~,~~~~~~~~~f = { t \over 2 } e^{ 2t } ~,~~~~~~~~e^{ 2 \phi - 2 \phi_0 } = 4 \sqrt{t\over 2}
 e^{ - t }~.
\ee

Let us now solve the equation in the region where it starts deviating from \nref{cvmn}.
We will call this the ``fivebrane'' region.
We can write $c = t + \mu(t)$ and we assume that $ \mu(t) \ll t$, but we do not assume that $\mu'$ is small.
The transition   happens around a value of $t $ we will call $t_5$, we will define it better below.
So we want $\mu \ll t_5$.
In equation \nref{equationftau} we set $\mu=0$ so that $f$ remains the same. In
 equation \nref{equationctau} we expand to first order in $\mu$
\be \label{muval}
\mu' = { 2 t \sinh^2 t \over f } \mu = { \mu } ~~~ \to  ~~~~ \mu = e^{t - t_5}
\ee
where $t_5$ is an integration constant.
When this is inserted in the expression for the metric and the  dilaton we obtain
\be \label{fivesol}
c' = 1 + e^{ t - t_5 } ~,~~~~~~~ e^{ 2 \phi - 2 \phi_0 } = 4 \sqrt{ t \over 2 } e^{-t} ( 1 + e^{ t - t_5 } )~.
\ee
We see that the transition in the behavior of $c'$ occurs at $t\sim t_5$ and it is very rapid compared to the
variation of $t$. Thus in the expression for the dilaton we can approximate the value of $t$ in the prefactor
as a constant equal to $t_5$. Therefore we can say that the dilaton changes from the large $t$ behavior in
\nref{cvmnlarge} to basically a constant.
It is possible to see that the full metric becomes that of an ordinary fivebrane if
we identify $ e^{ t - t_5 } = { r^2 \over M \alpha' } $. The directions along the $S^2$ are simply a constant.

 We get the following approximate metric
\bea
  { M \alpha' \over 4 }  ds^2_6 & \approx &  ( M \alpha' +  r^2 ) \left\{
  { 1 \over 4} dt^2 + { 1 \over 4 } [  (\epsilon_3 + A_3)^2 +e_1^2 + e_2^2 ] \right\}+
  {  M \alpha' t_5 \over 2 }
  ( \epsilon_1^2 + \epsilon_2^2 ) \notag
\\
 & \approx & ( 1 + { \alpha' M \over r^2 } ) ( dr^2 + r^2 d\tilde \Omega_3^2 ) + {  M \alpha' t_5 \over 2 } d\Omega_2^2
\\
  e^{ 2 \phi - 2 \phi_0 } & \approx &  4 \sqrt{ t_5 \over 2} e^{ t_5}  ( 1 + { \alpha' M \over r^2 } )
  \\
 &&  |t - t_5|  \ll \log t_5 ~,~~~~~~~~~~~  e^{ t -t_5} = {r^2 \over M \alpha' } \ll t_5 ~. \label{boundtf}
\eea
We see that $\tilde \Omega_3$ is a three sphere (fibered over the $S^2$). Since the $S^2$ is large compared
to the other dimensions, we can neglect the fact that the $S^3$ is fibered. This metric looks like the metric
of a fivebrane in flat space where $r$ and $\Omega_3$ are the four directions transverse to the fivebrane.
We have also specified the regime where the solution is valid. In this regime the size of the $S^2 $ is constant and very big.
 The upper bound on $t_5$ in \nref{boundtf}  comes from equating $t \sim \mu(t) \sim t_5$.
It is also convenient to estimate the value of $\gamma$ that we would obtain from this solution.
We do this
by simply extrapolating $\mu $ to the origin, where \nref{muval}   is  not really valid.
 That then gives the estimate
\be
\gamma^2 = c'(0) = 1 + e^{-t_5} ~,~~~~~~~~~~~~~~~~ \gamma^2 -1 \propto e^{ - t_5 }~.
\ee

We   now solve the equation in the region  where $t  \sim t_5$. In this region we   approximate
the function $k(\tau) \sim  t_5^2$. Then the energy  \nref{hamiltonian}
becomes conserved we get
\be
H = { 1 \over 16 }  { (\partial_\tau f )^2 \over \sqrt{f} } -  { k(\tau) \over \sqrt{f} }
=  e^{ 2 (\phi - \phi_0) } = E~.
\ee
We rewrite this in terms of $c$ to obtain
\bea
  ( c^2 - t_5^2) \de_\tau c  &= &E^2   ~,~~~~~~~~~~~~~ \tau = { e^{2t}  \over 8 }
\\
{ c^3 \over 3} - t_5^2 c &=& E^2 \tau  - { 2 \over 3 } t_5^3   \label{solla}
\\
E &=& e^{ 2 \phi_\infty - 2 \phi_0}
\eea
where we used the large $t$ limit of \nref{tauexp}. $E$ is an integration constant equal  to
the energy.
In addition we fixed another integration
constant by saying that $c = t_5$ for $\tau =0$. We can now determine the integration constant $E$
by matching to the previous expression \nref{fivesol}.
We expand \nref{solla} for small $\tau$ by writing $c = t_5 +   \mu$, as we did before.
We then find that
\be \label{matching}
  t_5   \mu^2 = E^2 \tau ~,~~~~~~~~~~~~~   \mu =
   { E    \over \sqrt{t_5 8 } } e^{ t}   = e^{ t -t_5} ~,~~~~
   E = \sqrt{ 8 t_5 } e^{-t_5}~.
   \ee
   where we solved for $\mu$ using \nref{solla} and equated it to our previous value \nref{muval}.
Once we have determined $E$, we can now determine the value of $t_\infty$ in this
solution by looking at the large $t$ behavior
\be
c \sim  { 1 \over 6 } e^{ 2 ( t - t_\infty ) \over 3 } \sim
 (  E^2   \tau 3  )^{1/3} = 3^{ 1/3}   t_5^{1/3} e^{ {2 \over 3} ( t-t_5)} ~.
\label{approxchat}
\ee
We see that
 \be \label{energtin}
 E \sim { e^{ - t_\infty} \over 9 }  ~~~{ \rm and} ~~~~~ e^{ - 2 t_\infty   } = ~3^{ 4} ~8 \, t_5 \, e^{ - 2 t_5  }~.
\ee
  We see that $t_5 = t_\infty + o (\log t_\infty)$, thus we can replace $t_5$ by $t_\infty$ in some  of the above formulas.

We can see that \nref{solla} reduces to the resolved conifold as follows. We introduce $\rho$ through
\be \label{cderho}
c = t_\infty + {\rho^2\over 6}
\ee
and we  write
\bea
\partial_t c &=&  { \rho^2 \over 9 } { \rho^2 + 18 t_\infty \over \rho^2 + 12 t_\infty } =
 { \rho^2  \over  9 } \kappa(\rho)
~,~~~~ \alpha^2 = 2 t_\infty \label{rcapp}
\\
\partial_t c  dt^2 & = & { \rho^2 d\rho^2 \over 9 \partial_t c } = {  d\rho^2 \over \kappa(\rho)  }
\\
&&  - \log t_\infty  \ll   \log {  \rho^2 \over \alpha^2 }   \ll  ( \log t_\infty )~. \label{rangeapp}
\eea
In this way we see that we recover the resolved conifold metric \nref{RCmetric}. We have also indicated the
range of $\rho$ were we can trust the resolved conifold metric.  In the lower bound we encounter the
near \horizon region of the fivebranes and in the upper bound we need to start taking into account the
``running'' of $\alpha^2$. The region of validity is very large for large $t_\infty$.

In summary, the solution has various regions. The transition between various regions happens at $t\sim t_\infty$, or
within  are region of size $\log t_\infty $ around this value.
For $t \ll t_\infty$ we have the CV-MN solution, which can be viewed as the near \horizon geometry of $M$ fivebranes.
 When $t\sim t_5$ we leave the near \horizon geometry and the dilaton becomes constant.
  For larger values of $t$, but still within $t/t_\infty $ of order one, we
 can view the solution as the resolved conifold
  plus some branes on the $S^2$.
   Notice that the metric behaves as the metric of the resolved conifold up to   $\rho^2 \sim 1$ at $t\sim  t_5$, see
   \nref{matching}, \nref{cderho}. This is much smaller than $ t_\infty$ which is setting the size of the sphere of the resolved
   conifold. Furthermore, the resolved conifold metric is accurate up to a value of $t$ where the ``running'' of the size starts
    to matter.
   In other words, the full metric \nref{ansatz} has a ratio of sizes of two-spheres going like $t$. Thus, we can approximate
   that ratio
   as a constant for values of $t$ such that $t/t_\infty$ is of order one. On the other hand, since the relation between
   $\rho$ and $t$ is
   exponential, we see that we can trust the metric of the conifold up to a value of $\rho$ such that $\log ( \rho/\alpha)
   \sim o(t_\infty) $, see \nref{rangeapp}.
   Thus, there is a large region of the geometry that is accurately given by the resolved conifold.
   For larger values of $t$, then we should take into account this ``running'', but by this stage, the resolution
   parameter is a small (but non-normalizable)
   deformation of the metric.

The above formulas  give us an accurate description of the solution in \nref{ansatz} before
we perform the boosting procedure.
However, in order to do the boosting, we will also need the behavior of the dilaton to
the next order. In order words, we want $ e^{ 2( \phi - \phi_\infty ) } -1 $.
It is convenient to write the expression for $f$ from \nref{solla} as
\be
f = { c^2 -t_\infty^2 \over \partial_\tau c}  ~,~~~~~~~~~ f^{1/2} =
 { E  \over  \partial_\tau c }
\ee

We can find the expression for the variation of the dilaton by
using the energy non-conservation equation \nref{ennoncon}
\bea
e^{ 2( \phi - \phi_\infty ) } -1  & = &
 e^{- 2(\phi_\infty -\phi_0)}\int_t^\infty  dt' { \partial_t h \over \sqrt{f} } =
 { 1 \over E^2 }
\int_t^\infty  dt' \partial_t h
  { \partial_\tau c  }\notag \\
& = &    { 2 \over E^2}  \int_t^\infty dt' \,  { t' \partial_\tau c  }~.
\eea
We now change variable from $t$ to $\rho$. We substitute
$c = t_\infty + {\rho^2/6}$ into \nref{solla}, obtaining
\be
e^{2t}~=~ \frac{1}{81 E^2} \rho^4 (\rho^2 + 18 t_\infty)~.
\ee
Then we have
\bea
\hat h =  e^{ 2( \phi - \phi_\infty ) } -1  & = &
 {  8 \over E^2} \int_{\rho}^\infty  d \rho' { \rho' \over 3  } t' e^{-2t'} \notag \\
& =  & 8 \int_{\rho}^\infty  d \rho { \rho\over 6  }\log\left[
{\rho^4 ( \rho^2 + 18 t_\infty) \over 81 E^2 } \right] { 81  \over  \rho^4 ( \rho^2 + 18 t_\infty ) }~.
\label{ptwarp}
 \eea
We see that the only remaining $E$ dependence is inside the log.
We can now estimate $\hat h$ in the range $\rho^2 \sim t_\infty$, but $ 1 \ll \rho^2$. The argument of the
logarithm in \nref{ptwarp} has a denominator that varies exponentially with $t_\infty$ \nref{energtin}.
Thus,  we can approximate
the logarithm  as $\log[~] \propto t_\infty$. Then all factors of $\rho$ outside the logarithm are approximated via
$\rho^2 \propto t_\infty $.  In this range we see that $\hat h \sim 1/t_\infty $.
Thus, $\hat h^{1/2} t_\infty \sim \sqrt{t_\infty}$
which justifies the approximation in \nref{swformula}.

In order to turn the asymptotics to precisely KS, we need to further redefine
coordinates, introducing a new coordinate $r$ defined via
\be
\rho = E^{1/3} r
\ee
Now we see that we get
\bea
\hat h = { 8 \over E^{4/3} }
\int_{r}^\infty  d r { r \over 6  }\log\left[
{r^4 ( r^2 + 18 \hat  t_\infty) \over 81 } \right] { 81  \over  r^4 ( r^2 + 18 \hat t_\infty ) }~,
~~~~~~~\hat t_\infty = t_\infty E^{ - 2/3}~.
 \eea
Now we have the standard expression for the KS asymptotics. Furthermore, this agrees with the
the expression of the warp factor of the solution by Pando-Zayas and Tseytlin \cite{PZT}, up to terms
that are important at small $t$ and probably  arise once we take into account the leading
order variation of the dilaton at small $t$, $t \sim t_5$.




\end{document}